\documentclass[prl,onecolumn,superscriptaddress,nofootinbib]{revtex4}
\usepackage[pdftex]{graphicx}
\usepackage{enumerate}
\usepackage{enumitem}
\usepackage{amsmath,amssymb,latexsym}
\usepackage{ascmac}
\usepackage{bm}
\newtheorem{theorem}{{\bf Theorem}}
\newtheorem{lemma}{{\bf Lemma}}
\def\U#1{{\rm #1}} 
\def\ph{\U{ph}}
\def\em{\U{em}}
\def\bit{\U{bit}}
\def\tr{\U{tr}}
\def\vac{{\rm vac}}

\newcommand{\code}{\U{code}}
\newcommand{\sample}{\U{sample}}
\newcommand{\wt}{\U{wt}}
\newcommand{\bra}[1]{\langle #1 |}
\newcommand{\ket}[1]{| #1 \rangle}
\newcommand{\expect}[1]{\left\langle #1 \right\rangle} 

\begin{document}
\title
{Differential-phase-shift QKD with practical Mach-Zehnder interferometer}
\author{Akihiro Mizutani}
\affiliation{Faculty of Engineering, University of Toyama, Gofuku 3190, Toyama 930-8555, Japan}
\author{Masanori Terashita}
\affiliation{Faculty of Engineering, University of Toyama, Gofuku 3190, Toyama 930-8555, Japan}
\author{Junya Matsubayashi}
\affiliation{Faculty of Engineering, University of Toyama, Gofuku 3190, Toyama 930-8555, Japan}
\author{Shogo Mori}
\affiliation{Faculty of Engineering, University of Toyama, Gofuku 3190, Toyama 930-8555, Japan}
\author{Ibuki Matsukura}
\affiliation{Faculty of Engineering, University of Toyama, Gofuku 3190, Toyama 930-8555, Japan}
\author{Suzuna Tagawa}
\affiliation{Faculty of Engineering, University of Toyama, Gofuku 3190, Toyama 930-8555, Japan}
\author{Kiyoshi Tamaki}
\affiliation{Faculty of Engineering, University of Toyama, Gofuku 3190, Toyama 930-8555, Japan}

\begin{abstract}
  {Differential-phase-shift (DPS) quantum key distribution stands as a promising protocol due to its simple 
  implementation, which can be realized with a train of coherent pulses and a passive measurement unit. 
To implement the DPS protocol, it is crucial to establish security proofs incorporating practical imperfections in users' devices, however, existing security proofs make unrealistic assumptions on the measurement unit using a Mach-Zehnder interferometer. In this paper, we enhance the implementation security of the DPS protocol by incorporating a major imperfection in the measurement unit. Specifically, our proof enables us to use practical beam splitters with a 
known range of the transmittance rather than the one with exactly 50\%, as was 
assumed in the existing security proofs. 
Our numerical simulations demonstrate that even with fluctuations of $\pm0.5\%$ in the transmittance from the ideal value, 
  the key rate degrades only by a factor of 0.57. 
  This result highlights the feasibility of the DPS protocol with practical measurement setups.
} 
 \end{abstract}

\maketitle

\section{introduction}
Quantum key distribution (QKD)~\cite{tamaki2014} 
enables distant parties to achieve information-theoretically secure communication. 
Among major QKD protocols~\cite{ekert91,bennett92,brub98,sarg,cow,continuous,inoue2002,R21,R22}, 
the differential-phase-shift (DPS) QKD protocol~\cite{inoue2002} has a feature with its simple implementation, involving 
a train of coherent pulses from a laser source and a passive measurement unit. 
Due to its simplicity, several experiments were conducted in Refs.~\cite{ex00,ex0,ex1,tokyoqkd,citemust}, 
including field demonstration in the Tokyo QKD network~\cite{tokyoqkd}. 
On the other hand, contrary to the simplicity in the experiments, the security proof of this protocol was a challenging problem. 
The difficulty arises from the fact that the secret key is extracted from the relative phases of adjacent pulses and all the pulses are interconnected, leading to the necessity of considering a large Hilbert space 
by taking the tensor product of Hilbert spaces of all the emitted pulses.

To overcome this difficulty, 
previous information-theoretic security proofs~\cite{dps2009,dps2012,dps2017,dps2019,dps2020,dps2023,dps2024} 
introduced blocks comprising several emitted pulses and considered extracting at most one-bit secret key from each block. 
This is also the case for DPS type protocols
\footnote{
Note that the DPS protocol is categorized as distributed-phase-reference QKD, and another prominent protocol is the coherent-one-way (COW) protocol~\cite{cow1,cow2,cow3,cow4,cow5}.}, 
such as the round-robin DPS protocol~\cite{rrdps,rrdpsmi,rrdpsex,rrdpssa,akgo,rrdpsnatp,rrdpsnatc}
\footnote{
Inspired by the round-robin DPS protocol, the Chau15 protocol~\cite{chau} was proposed, which is a qudit-based protocol and has a high bit-error tolerance, and its proof-of-principle demonstration was executed in Ref.~\cite{must2}.}, 
the small-number-random DPS protocol~\cite{hatake2017} 
and the differential quadrature phase shift protocol~\cite{dqps}. 
In particular, Ref.~\cite{dps2020} provides a security proof under the most relaxed assumptions for the source 
device, revealing that as long as the source emits identical and independent states, 
the security of the DPS protocol can be guaranteed. 
Interestingly, 
this work does not assume exact knowledge about the emitted states; the amount of privacy amplification can be 
determined according to statistics that Alice obtains from the source characterization experiment in which she measures 
the photon number distribution up to three photons. 
Although 
the assumptions on the source devices in the DPS protocol were relaxed so far, 
all the existing security proofs~\cite{dps2009,dps2012,dps2017,dps2019,dps2020,dps2023,dps2024,endo22,arxivsand} 
made ideal assumptions on Bob's measurement unit
\footnote{
Note that in the recent work on fully-passive QKD~\cite{fpassive,must3}, one-bit and two-bit delay Mach-Zehnder interferometers are used to respectively encode decoy states and key bits. 
In Refs.~\cite{fpassive,must3}, they assume perfect Mach-Zehnder interferometers with BSs having ideal transmittance.
}; 
the transmittance of the beam splitters (BSs) inside Bob's 
Mach-Zehnder interferometer is assumed to be exactly 50\%. Unfortunately, however, such an assumption is demanding because it is almost unfeasible to manufacture the perfect BS in practice. 
To implement the DPS protocol in the real world, it is crucial to establish security proofs that take into account imperfections in the measurement device.

In this paper, we relax this demanding assumption to employ a more feasible BS in which the transmittance surely lies within a certain range. Based on this more experimentally friendly assumption, we provide an information-theoretic security proof of the DPS protocol. 
Our security proof is based on complementarity, i.e., phase error correction approach~\cite{koashi2009}. 
In this approach, we consider an entanglement-based QKD protocol, where Alice virtually prepares entangled states between qubits and emitted states to Bob. 
In this virtual protocol, Alice extracts her sifted key by measuring her qubits in the key generation basis after Bob makes announcements about which pulse results in a successful detection. 
The crux of the complementarity approach is to derive the number of phase error events, where Alice fails the prediction of the outcome if her qubit were measured in the complementary basis to the key generation basis. Once the upper bound on the phase error rate is obtained, it is straightforward to determine the amount of privacy amplification from Theorem in Ref.~\cite{koashi2009}. 
Importantly, our security proof does not need the relativistic constraint to satisfy the sequential assumption, where Alice must wait to emit the next pulse until Bob completes the detection of the previous pulses. 
This assumption is needed to prove the security of the DPS protocol based on the entropy accumulation 
technique~\cite{arxivsand}. 

As a result of our security proof, we numerically simulate the resulting key rate (see Fig.~\ref{fig:kerate}), 
and it demonstrates that even under practical fluctuations $\pm$0.5\% and $\pm$1\% in 
the transmittance from the ideal value, 
the respective key rates are found to degrade only by a factor of 0.57 and 0.27. 
This result shows that the key rate does not degrade drastically 
even under practical fluctuations in the transmittance of the BSs, 
which suggests the feasibility of the DPS protocol with realistic measurement setups. 

The rest of the paper is organized as follows. 
First, in Sec.~\ref{sec:depro}, we explain the DPS protocol including the assumptions on users' devices. Next, in Sec.~\ref{sec:secru} we prove the security of our DPS protocol 
based on complementarity~\cite{koashi2009}. After that, in Sec.~\ref{sec:sim} 
we present our simulation results of the DPS protocol and compare the key rates 
assuming different ranges of the transmittance: 50\%$\pm$0\%, 50\%$\pm$0.5\%, and 50\%$\pm$1\%. 
Finally, we summarize the paper in Sec.~\ref{sec:conc}.

\section{DPS QKD with practical Mach-Zehnder interferometer}
\label{sec:depro}

\begin{table}[h]
\begin{center}
\begin{tabular}
{|l|l|}\hline 
Random variables and sets &Definition\\ \hline\hline
$b_i$ & Alice's bit randomly chosen for the $i$th emitted pulse in the block with $i\in\{1,2,3\}$ 
\\\hline
$\bm{b}$ &Abbreviation of $b_1b_2b_3\in\{0,1\}^3$
\\\hline
$\mathcal{S}$ & Index set of the detection events\\\hline
$N_{\rm det}$ & Cardinality of set $\mathcal{S}$ of the detection events\\\hline
${\rm TS}j_i$ 
& Time slot of detection of the $j_i$th ($j_i\in\{1,2\})$ pulse pair for the $i$th ($i\in\mathcal{S}$) detection event\\\hline
$d_i$ &Bob's raw key bit representing which of the detectors clicks for the $i$th ($i\in\mathcal{S}$) detection event \\\hline
$\mathcal{S}_{\rm code}$ & Index set of the detected and code events \\\hline
$N_{\rm code}$ & Cardinality of set $\mathcal{S}_{\rm code}$\\\hline
$\mathcal{S}_{\rm sample}$ & Index set of the detected and sample events \\\hline
$N_{\rm sample}$ & Cardinality of set $\mathcal{S}_{\rm sample}$\\\hline
$\kappa_B$ & Bob's sifted key $(d_i)_{i\in {\mathcal{S}_{\rm code}}}\in\{0,1\}^{N_{\rm code}}$ \\\hline
$\kappa^{\rm sample}_B$ & Bob's sample bit sequence $(d_i)_{i\in {\mathcal{S}_{\rm sample}}}\in\{0,1\}^{N_{\rm sample}}$ 
\\\hline
$\kappa_A$ & Alice's sifted key $(b_{j_i}\oplus b_{j_i+1})_{i\in \mathcal{S}_{\rm code}}\in\{0,1\}^{N_{\rm code}}$\\\hline
$\kappa^{\rm sample}_A$ & Alice's sample bit sequence $(b_{j_i}\oplus b_{j_i+1})_{i\in \mathcal{S}_{\rm sample}}
\in\{0,1\}^{N_{\rm sample}}$ \\\hline
$N_{\rm EC}$ & Length of the pre-shared secret key  consumed for bit error correction\\\hline
$\kappa^{\rm rec}_B$ & Bob's $N_{\rm code}$-bit reconciled key after bit error correction\\\hline
$N_{\rm PA}$ & Length of the bits discarded in privacy amplification\\\hline
$\ell$ &Net growth of the secret key we obtain when our DPS protocol is executed\\\hline
$Q$ & 
Ratio $N_{\rm det}/N_{\rm em}$ of the number of detected events to the total number of emitted blocks
\\\hline
$e_{\rm bit}$ & Bit error rate between Alice and Bob's sample bit sequences $\kappa^{\rm sample}_A$ and 
$\kappa^{\rm sample}_B$\\\hline
      \end{tabular}
\end{center}
\caption{A list of random variables and sets used throughout this paper}
 \label{tableI}
\end{table}

\begin{table}[h]
\begin{center}
\begin{tabular}
{|l|l|}\hline 
Symbols, systems&Definition\\ 
and states  &\\ \hline\hline
$N_{\rm em}$ & Number of total blocks sent by Alice\\ \hline
$S_i$ & Alice's quantum system of the $i$th emitted pulse in the block with $i\in\{1,2,3\}$\\\hline
$\bm{S}$ & Abbreviation of $S_1S_2S_3$\\ \hline
$\hat{\rho}^{b_i}_{S_i}$ & Quantum state of the $i$th emitted pulse in the block with $i\in\{1,2,3\}$\\\hline
$\hat{\rho}^{\bm{b}}_{\bm{S}}$ & Quantum state of a single emitted block of systems $\bm{S}$, that is, 
$\bigotimes_{i=1}^3\hat{\rho}_{S_i}^{b_i}$\\\hline
$R_i$ & Alice's quantum system purifying $\hat{\rho}^{b_i}_{S_i}$, which we assume Eve has no access to\\ \hline
$\ket{\psi_{b_i}}_{S_iR_i}$ & Purified state of $\hat{\rho}_{S_i}^{b_i}$\\ \hline
$q_n$ & Upper bound on the probability of single block $\hat{\rho}^{\bm{b}}_{\bm{S}}$ 
emitting $n$ or more photons\\ \hline
$\eta_{\rm det}$ & Quantum efficiencies of Bob's detectors, which are assumed to be identical for all the 
detectors\\ \hline
$\eta_1$ and $\eta_2$ & Transmittance of two beam splitters (BSs) in Bob's Mach-Zehnder interferometer\\ \hline
$\mathcal{R}_k$ & Range of the transmittance of the $k$th BS, 
$[1/2-\delta^{(\rm BS)}_k, 1/2+\delta^{(\rm BS)}_k]$, where $k\in\{1,2\}$\\
&and $0\le\delta^{(\rm BS)}_k<0.5$. We assume that the actual transmittance lies within this range. 
\\\hline
$\eta^U_k$ and $\eta^L_k$ & Upper and lower bounds on the transmittance of the $k$th BS with $k\in\{1,2\}$,
\\ 
&namely, $1/2+\delta^{(\rm BS)}_k$ and $1/2-\delta^{(\rm BS)}_k$, respectively\\\hline
$(l,i)$ & Label of the $i$th pulse received by Bob passing through the lower arm of \\
&the Mach-Zehnder interferometer with $i\in\{1,2,3\}$ \\ \hline
$(u,i)$ & Label of the $i$th pulse received by Bob passing through the upper arm of \\
&the Mach-Zehnder interferometer with $i\in\{1,2,3\}$ \\ \hline
$t$ & Probability of Bob choosing the code event for each detected event
\\\hline
      \end{tabular}
\end{center}
\caption{A list of symbols, quantum systems and states used throughout this paper}
 \label{tableII}
\end{table}

In this section, we explain our assumptions on Alice and Bob's devices and describe our DPS protocol. 
To facilitate the reader's understanding, we summarize in Table~\ref{tableI} the definitions of the random variables and sets, and in Table~\ref{tableII}, the definitions of quantum systems, states, and symbols that appear in the assumptions on the devices and our DPS protocol.

\subsection{Assumptions on devices}
\label{sec:deviceass}
For Alice's source device, we assume the following conditions. 
\begin{enumerate}[label=(A\arabic*)]
\item
For each pulse emission, Alice uniformly and randomly chooses bit $b_i\in\{0,1\}$, and according to the chosen bit, 
she prepares state $\hat{\rho}_{S_i}^{b_i}$ of system $S_i$. We call consecutive three emitted pulses block, and the state of a single block is written as
\begin{align}
\hat{\rho}_{\bm{S}}^{\bm{b}}:=\bigotimes_{i=1}^3\hat{\rho}_{S_i}^{b_i}
\end{align}
with $\bm{S}:=S_1S_2S_3$ and $\bm{b}:=b_1b_2b_3$. 
Here, := is the mathematical symbol to define the left-hand side by the right-hand side. 
We suppose that bit information $b_i$ is only encoded to the $i$th emitted pulse, and Eve cannot access to system $R_i$ that purifies state 
$\hat{\rho}_{S_i}^{b_i}$. $\ket{\psi_{b_i}}_{S_iR_i}$ denotes the purified state of $\hat{\rho}_{S_i}^{b_i}$. 
\item
The probabilities of the $i$th emitted pulse being the vacuum state are independent of the chosen bit, namely, 
\begin{align}
\tr(\hat{\rho}_{S_i}^0\ket{\vac}\bra{\vac})=\tr(\hat{\rho}_{S_i}^1\ket{\vac}\bra{\vac}).
\label{eq:vace}
\end{align}
Here, $\ket{\vac}$ denotes the vacuum state. 
\item
We assume that the probability of any block emitting $n$ or more photons 
is upper-bounded by $q_n$ for $n\in\{1,2,3\}$, i.e., 
\begin{align}
\sum_{m\ge n}\tr(\hat{\rho}_{\bm{S}}^{\bm{b}}\ket{m}\bra{m})\le q_n.
\label{def:qn}
\end{align} 
Here, $\ket{m}$ denotes the $m$ photon-number state.
\end{enumerate}
As for Bob's measurement unit, we assume the following conditions. 
\begin{enumerate}[label=(B\arabic*)]
\item
\label{ass:b2}
Bob employs two photon-number-resolving (PNR) detectors $D_0$ and $D_1$ 
that discriminate between the vacuum, a single photon, and two or more photons of a specific single optical mode. 
The detection inefficiency is modeled as a beam splitter (BS) followed by an ideal detector with a unit quantum efficiency. 
The quantum efficiencies are identical for both PNR detectors and are denoted by $\eta_{\rm det}$. 
Moreover, we assume that the dark counting of the detector is simulated by a stray photon source positioned 
in front of Bob's measurement unit. 
\item
\label{ass:b1}
Let $\eta_1$ and $\eta_2$ be transmittance of two BSs in the Mach-Zehnder interferometer 
with respect to the single optical mode detected by the detectors. 
For later convenience, a BS with transmittance $\eta$ is denoted by $\eta$-BS. 
The transmittance of the BSs is assumed to be constant during the execution of the QKD protocol. 
Alice and Bob do not know the exact transmittance but its ranges:
\begin{align}
\eta_1\in \mathcal{R}_1:=[1/2-\delta^{(\U{BS})}_1,1/2+\delta^{(\U{BS})}_1]~\U{and}~
\eta_2\in  \mathcal{R}_2:=[1/2-\delta^{(\U{BS})}_2,1/2+\delta^{(\U{BS})}_2]
\label{eq:symmetricrange}
\end{align}
with $0\le\delta^{(\U{BS})}_1,\delta^{(\U{BS})}_2<1/2$. For simplicity of notations, we define 
\begin{align}
\eta^{L}_i=1/2-\delta^{(\U{BS})}_i~\U{and}~\eta^{U}_i=1/2+\delta^{(\U{BS})}_i.
\label{eq:etabound}
\end{align}
We discuss the practicality of this assumption~\ref{ass:b1} as follows. 
In the actual manufacturing process of BSs, manufactures aim to achieve an ideal transmittance of 50\%. 
However, achieving this exact ideal transmittance is almost impossible in practice, and it is inevitable that the manufactured BSs have fluctuations in transmittance around 50\%. 
If the range of fluctuations [i.e., $\delta^{(BS)}_1$ and $\delta^{(BS)}_2$ in Eq.~(\ref{eq:symmetricrange})] 
is set conservatively large enough, then Eq.~(\ref{eq:symmetricrange}) could be satisfied with real-world BSs. 
Consequently, practical BSs can be securely employed in executing our DPS protocol.
\end{enumerate}

For simplicity of our security proof in Sec.~\ref{sec:secru}, 
we assume in~\ref{ass:b2} that Bob employs PNR detectors that 
can discriminate between the vacuum, a single photon, and two or more photons. 
However, even when Bob employs threshold detectors that only distinguish whether photons arrived or not, 
the security of our DPS protocol can be guaranteed. 
We discuss the security with threshold detectors in Appendix~\ref{app:thde}.

\subsection{Protocol description}
\label{sec:proto}
\begin{figure*}[t]
\includegraphics[width=11cm]{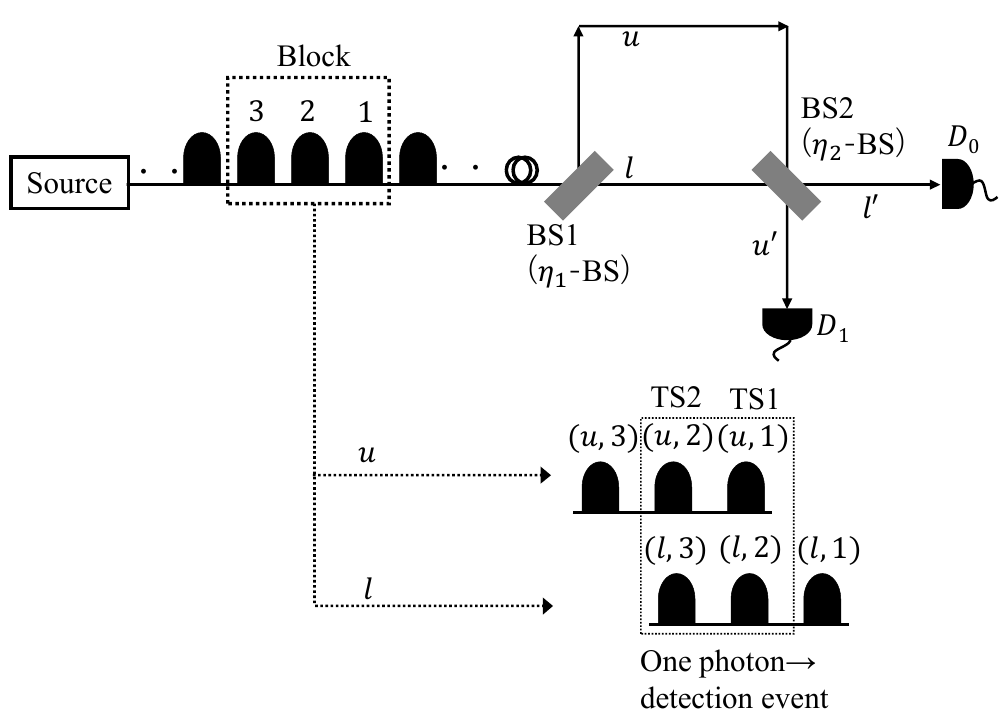}
\caption{
Experimental setup for our DPS protocol. Alice sends blocks composed 
of three pulses to Bob, who receives them with the one-bit
delay Mach-Zehnder interferometer and PNR detectors $D_0$ and $D_1$. 
The difference between our work and the previous ones~\cite{dps2019,dps2020,dps2023} lies in the transmittance 
$\eta_1$ and $\eta_2$ of Bob's beam splitters, which are not necessarily 50\% as previously assumed, but 
$\eta_1$ and $\eta_2$ can respectively take any value within the ranges $\mathcal{R}_1$ and $\mathcal{R}_2$. 
$u$ and $l$ respectively represent the upper and lower arms of the Mach-Zehnder interferometer, and 
$u'$ and $l'$ denote the output modes of the second BS (BS2). 
The pulse pairs $(u,1)$ and $(l,2)$, and $(u,2)$ and $(l,3)$ interfere at the BS2, and TS1 (TS2) 
is the time slot of detection of the first (second) pulse pair. 
A detection event occurs when Bob detects a single photon in total among the time slots TS1 and TS2. 
}
 \label{fig:actual}
\end{figure*}
Here, we describe the procedures of our DPS protocol (see Fig.~\ref{fig:actual}). 
Our protocol is identical to the previous works~\cite{dps2019,dps2020,dps2023} 
with the only difference being the transmittance of the beam splitters (BSs) inside Bob's measurement unit.

\begin{enumerate}
\item
Alice and Bob respectively execute the following steps (a) and (b) $N_{\em}$ times. 
\begin{enumerate}
\item
Alice uniformly and randomly chooses three bits $\bm{b}=b_1b_2b_3\in\{0,1\}^3$, and according to the chosen bits, 
she sends state 
$\hat{\rho}_{\bm{S}}^{\bm{b}}$ of a single block to Bob via a quantum channel. 
\item
Bob splits the incoming three pulses into two pulse trains using the first BS (BS1). 
The $i$th pulse with $i\in\{1,2,3\}$ passing through the lower and upper arms of the Mach-Zehnder interferometer 
are labeled by $(l,i)$ and $(u,i)$, respectively. 
The pulse pairs $(u,1)$ and $(l,2)$, and $(u,2)$ and $(l,3)$ interfere at the second BS (BS2). 
We define the time slots of detection of the first and second pulse pairs as TS1 and TS2, respectively. 
We define a ``detection event" as the one in which Bob detects one photon in total in TS1 and TS2. 
The detection 
event at TS$j$ (with $j\in\{1,2\}$) determines the raw key bit $d$ depending on which of the two detectors clicks.
\end{enumerate}
\item
Bob defines the set of detection events $\mathcal{S}\subset \{1,...,N_{\em}\}$ with length 
$|\mathcal{S}|:=N_{\det}$, the set of time slots at which Bob obtained the detection event, 
 i.e., $\{\U{TS}j_i\}_{i\in\mathcal{S}}$, and the raw key bits $\bm{d}:=(d_i)_{i\in\mathcal{S}}$. 
Here, $j_i$ and $d_i$ ($i\in\mathcal{S}$) respectively denote the values of $j$ and $d$ of the $i$th detection event. 
Within the detection events, 
Bob randomly assigns each detection event to a code event with probability $t$ or a sample event with probability 
$1-t$ (where $0<t<1$). 
Then, he obtains the code set $\mathcal{S}_{\code}$ with length $|\mathcal{S}_\code|:=N_{\code}$, 
the sample set $\mathcal{S}_{\sample}$ with length $|\mathcal{S}_\code|:=N_{\code}$, 
his sifted key $\kappa_B:=(d_i)_{i\in\mathcal{S}_{\code}}$, and the sample bit sequence 
$\kappa^{\sample}_B:=(d_i)_{i\in\mathcal{S}_{\sample}}$. 
\item
Bob announces $\mathcal{S}_{\code}, \mathcal{S}_{\sample}$, $\{\U{TS}j_i\}_{i\in\mathcal{S}}$ and 
$\kappa_B^\sample$ via an authenticated public channel.
\item
Alice obtains her sifted key $\kappa_A:=(b_{j_i}\oplus b_{j_i+1})_{i\in\mathcal{S}_\code}$ 
and the sample bit sequence $\kappa^{\sample}_A:=(b_{j_i}\oplus b_{j_i+1})_{i\in\mathcal{S}_\sample}$. 
\item
Alice estimates the bit error rate in the code events from the bit error rate in the sample events, selects a bit error correction code, and sends the syndrome information of her sifted key $\kappa_A$ to Bob by consuming pre-shared secret key 
of length $N_{\U{EC}}$. 
Using the syndrome information, Bob corrects bit errors in his sifted key and obtains the reconciled key 
$\kappa^{\rm rec}_B$. 
\item
Alice and Bob execute privacy amplification to respectively shorten $\kappa_A$ and $\kappa_B^{\U{rec}}$ by 
$N_\U{PA}$ to obtain their final keys of length $N_{\code}-N_\U{PA}$. 
\end{enumerate}
After the execution of the protocol, the net length of the increased secret key is given by
\begin{align}
\ell=N_{\code}-N_{\U{PA}}-N_{\U{EC}}.
\end{align}
For later use, we define the following parameters
\begin{align}
Q:=\frac{N_{\det}}{N_{\em}},~e_{\bit}:=\frac{\wt(\kappa_A^{\sample}\oplus\kappa_B^{\sample})}{N_{\sample}},
\label{eq:exdef}
\end{align}
where wt$(a)$ represents the weight, i.e., the number of ones in the bit sequence $a$. 

\section{security proof}
\label{sec:secru}

\begin{table}[h]
\begin{center}
\begin{tabular}
{|l|l|l|}\hline 
Systems, states&Definition & Reference\\ 
and operators & & \\ \hline\hline
$A_i$ & Alice's fictitious qubit initially entangled with the $i$th emitted state in a block &Eq.~(\ref{fekj})
\\
&with $i\in\{1,2,3\}$. This qubit remains at Alice's laboratory during the protocol. & 
\\\hline
$\bm{A}$ & Abbreviation of $A_1A_2A_3$ & Eq.~(\ref{fekj})
\\\hline
$\ket{1}_B$ and $\ket{3}_B$ & Single photon state in the pulse $(u,1)$ and $(l,3)$, respectively
&Eqs.~(\ref{ertjklghertjgyh})-(\ref{bobpovm4})
\\\hline
$\ket{2}_B$ & Single photon state in the second pulse incoming to the first BS. Note that&
Eqs.~(\ref{ertjklghertjgyh})-(\ref{bobpovm4})\\
&$\ket{2}$ and $\ket{3}$ are {\it not} two-photon and three-photon states but single-photon states. &\\\hline
$B$ & Bob's system indicating the Hilbert space spanned by $\ket{1}_B, \ket{2}_B$ and $\ket{3}_B$&Eqs.~(\ref{ertjklghertjgyh})-(\ref{bobpovm4}) \\\hline
$\hat{\Pi}_{j,D_b}$ & POVM element for a detection event in detector $D_b$ at time slot TS$j$ & Eqs.~(\ref{ertjklghertjgyh})-(\ref{bobpovm4})\\
& with $b\in\{0,1\}$ and $j\in\{1,2\}$ & 
\\\hline
$\hat{e}^{(\U{TS}j)}_{\bit}(\eta_1,\eta_2)$ & POVM element for a bit error event at time slot TS$j$ with $j\in\{1,2\}$
& Eq.~(\ref{eq:laterebit})\\\hline
$\hat{e}_{\bit}(\eta_1,\eta_2)$ & POVM element for a bit error event & Eq.~(\ref{koret1}) \\\hline
$\hat{e}^{(\U{TS}j)}_{\ph}(\eta_1)$ & POVM element for a phase error event at time slot TS$j$ with $j\in\{1,2\}$
& Eqs.~(\ref{eq:ephpovmts1}), (\ref{eq:ephpovmts2})\\\hline
$\hat{e}_{\ph}(\eta_1)$ & POVM element for a phase error event
& Eq.~(\ref{koret2})\\\hline
$\hat{P}_a$ & Projector acting on Alice's qubits $\bm{A}$, projecting onto the subspace
& Eqs.~(\ref{eq:defP1}), (\ref{eq:exp3}),\\
& with the weight of $a$ along the  $Z$-basis with $a\in\{0,1,2,3\}$ & (\ref{eq:exp0}) and (\ref{eq:exp2}) \\\hline
      \end{tabular}
\end{center}
\caption{A list of quantum systems, states, and operators}
 \label{tableIII}
\end{table}

\begin{table}[h]
\begin{center}
\begin{tabular}
{|l|l|l|}\hline 
Random variables&Definition & Reference\\ 
and functions & & \\ \hline\hline
$z_{A_j}\in\{0,1\}$ & Measurement outcome if qubit $A_j$ were measured in the $Z$-basis with $j\in\{1,2\}$.
 &Fig.~\ref{fig:virbob}
\\
&In our complementarity security proof, Alice's task is to predict $z_{A_j}$.
& 
\\\hline
$e_{\rm ph}$  & Ratio of the number of phase errors (wrong predictions of $z_{A_j}$) to $N_{\rm code}$ &
Ths.~\ref{th:keyrate}, \ref{theorem2}
\\\hline
$e^U_{\rm ph}$  & Upper bound on $e_{\rm ph}$ & Eq.~(\ref{def:theo2})\\\hline
$\lambda(\eta_1,\eta_2)$ & Function that appears in the expression of $e^U_{\rm ph}$ &Eq.~(\ref{def:lambdade})\\\hline
$t_{\rm Bob}$ & Bob's hint to help Alice predict $z_{A_j}$ &Fig.~\ref{fig:virbob} 
\\\hline
$\omega_i\in\Omega$ & $i$th measurement outcome in the virtual measurement on systems $\bm{A}B$ 
to derive $e^U_{\rm ph}$  & Eq.~(\ref{eq:Omega}) \\\hline
$\chi^{(i)}_{\rm ph}\in\{0,1\}$ & $i$th random variable taking the value 1 if Alice and Bob obtain a phase error& 
Eq.~(\ref{eq:cph})\\
($i\in\{1,...,N_{\det}\}$) & from the $i$th measurement; otherwise, it takes the value 0&
\\\hline
$\chi^{(i)}_{\rm a}\in\{0,1\}$ & $i$th random variable taking the value 1 if Alice and Bob obtain weight $a$ 
& 
Eq.~(\ref{eq:ca})\\
($i\in\{1,...,N_{\det}\}$) & from the $i$th measurement of $\{\hat P\}_a$; otherwise, it takes the value 0&
\\\hline
$\chi^{(i)}_{\rm bit}\in\{0,1\}$ & $i$th random variable taking the value 1 if Alice and Bob obtain a bit error& 
Eq.~(\ref{eq:cbit})\\
($i\in\{1,...,N_{\det}\}$) & from the $i$th measurement; otherwise, it takes the value 0&
\\\hline
$N_{\xi}$ & Random variable that can be obtained by summing $\chi^{(i)}_{\xi}$ over the entire number of 
& Eq.~(\ref{eq:Nxi})\\
with $\xi\in\{{\rm ph, bit}, a\}$ & probabilistic trials, namely, $\sum_{i=1}^{N_{\rm det}}\chi^{(i)}_\xi$
 &
\\\hline
$X^{(i)}_{\xi}$ ($i\in\{1,...,N_{\det}\}$) 
& $i$th random variable to apply Azuma's inequality  & 
Eq.~(\ref{eq:rX})\\
with $\xi\in\{{\rm ph, bit}, a\}$ &to relate the random variable $N_{\xi}$ and its expectation &
\\\hline
\end{tabular}
\end{center}
\caption{A list of random variables and functions}
 \label{tableIV}
\end{table}

In this section, we present the security proof of our DPS protocol. In Sec.~\ref{bob:pobmdp2}, 
we introduce virtual procedures conducted by Alice and Bob. 
When evaluating the security of the sifted key based on complementarity~\cite{koashi2009}, 
we are interested in how accurately Alice can predict the outcome of the measurement that is complementary to the one for obtaining the sifted key, and the virtual protocol is useful to consider this scenario. 
As the parameter to quantify the accuracy of the prediction, we employ the phase error rate, 
which determines the amount of privacy amplification, 
and those errors are events in which Alice fails to predict the complementary measurement outcomes. 
In Sec.~\ref{bob:comple}, we discuss the relationship between the number of phase errors and the amount of privacy amplification performed in the actual protocol. 
Phase errors cannot be directly observed in the actual protocol, and instead they have to be estimated from the quantities that can be observed in the actual experiment. For this, 
Sec.~\ref{bob:povms} introduces the operators for obtaining bit and phase error events, and then 
in Sec.~\ref{bob:derivation}, we derive the upper bound on the number of phase errors using experimentally observed data. 

To help the reader, we summarize in Table~\ref{tableIII} the definitions of quantum systems, states, and operators, 
and in Table~\ref{tableIV}, the definitions of random variables and functions that appear in this Sec.~\ref{sec:secru}.

\subsection{Alternative procedures for Alice and Bob}
\label{bob:pobmdp2}
\begin{figure*}[t]
\includegraphics[width=3.5cm]{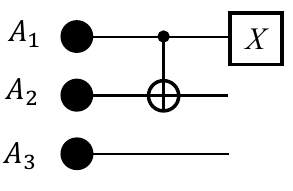}
\caption{
Alice's operation in the virtual scheme to obtain her sifted key bit when a detection event occurs at time slot TS1. 
She inputs the first and second qubits $A_1$ and $A_2$ to the C-NOT gate followed by measuring the first qubit in the $X$-basis to obtain her sifted key bit. 
}
 \label{fig:viralice}
\end{figure*}
In the security proof, it is convenient to consider the virtual protocol in which Alice prepares the following entangled state 
\begin{align}
\ket{\Phi}_{\bm{ASR}}:=\bigotimes_{i=1}^3
\sum_{b_i=0}^1
\frac{\hat{H}\ket{b_i}_{A_i}\ket{\psi_{b_i}}_{S_iR_i}}{\sqrt{2}},
\label{fekj}
\end{align}
keeps qubits $\bm{A}:=A_1A_2A_3$ and sends system $\bm{S}:=S_1S_2S_3$ to Bob. 
Here, $\hat{H}$ is the Hadamard operator, and the $Z$-basis for the $j$th qubit are defined by $\{\ket{0}_{A_j},\ket{1}_{A_j}\}$. 
In this virtual scheme, Alice's sifted key when a detection event occurs at time slot TS$j$ (with $j\in\{1,2\}$) 
is obtained by applying the controlled-not (C-NOT) gate with system 
$A_j$ ($A_{j+1}$) being the control (target) qubit followed by measuring system $A_j$ 
in the $X$-basis (see Fig.~\ref{fig:viralice}). 
Here, the $X$-basis states are defined by $\{\ket{+},\ket{-}\}$ with $\ket{\pm}:=(\ket{0}\pm\ket{1})/\sqrt{2}$. 
Importantly, accessible quantum information to Eve is the same as the one of the actual protocol, 
and therefore we can employ this virtual scenario for the security analysis.

Regarding Bob's measurement, we 
consider a measurement setup where two $\eta_{\det}$-BSs representing the quantum efficiency of 
the detectors are placed in front of Bob's interferometer. The following theorem guarantees the equivalence between such a measurement and the actual measurement. 
The proof of Theorem~\ref{theorem:bsposition} is provided in Appendix~\ref{app:pT}. 
\begin{theorem}
\label{theorem:bsposition}
Under the assumptions \ref{ass:b2} and \ref{ass:b1} on detectors in Sec.~\ref{sec:deviceass}, 
the beam splitters (BSs) representing the quantum inefficiency of the detectors can be forwarded in front of Bob's Mach-Zehnder interferometer.
\end{theorem}
When a detection event occurs, Bob's state in system $B$ can be expressed with 
the orthogonal basis $\mathcal{B}:=\{\ket{1}_B,\ket{2}_B,\ket{3}_B\}$. 
Here, $\ket{1}$ $(\ket{3})$ represents that a photon exists in the pulse $(u,1)$ (pulse $(l,3)$), 
while $\ket{2}$ indicates that the single photon exists in the pulse $(l,2)$ or $(u,2)$, namely, 
the second incoming pulse to the BS1 (see Fig.~\ref{fig:actual}). 
The measurement operator for detecting one photon in detector $D_i$ at time TS$j$ with $i\in\{0,1\}$ and $j\in\{1,2\}$ 
is expressed as $\hat{P}(\hat{a}^\dagger_{j,D_i}\ket{\vac})$. 
The projector $\hat{P}(\ket{\cdot})$ is defined as $\ket{\cdot}\bra{\cdot}$, and $\hat{a}^\dagger_{j,D_i}$ 
denotes the creation operator for a photon just before detector $D_i$ at time slot TS$j$. 
By taking the time reverse of the BS1 and BS2 such that $\hat{P}(\hat{a}^\dagger_{j,D_i}\ket{\vac})$ 
is written with respect to $\mathcal{B}$, 
we obtain the following POVM (positive operator valued measure) elements $\hat{\Pi}_{j,D_i}$ 
for obtaining a detection event in detector $D_i$ at time TS$j$. 
For completeness of the paper, we give their derivations in Appendix~\ref{app:devMP}. 
\begin{align}
\hat{\Pi}_{j=1,D_0}=&\hat{P}(\sqrt{1-\eta_2}\ket{1}_B+\sqrt{\eta_1\eta_2}\ket{2}_B),
\label{ertjklghertjgyh}
\\
\hat{\Pi}_{j=1,D_1}=&\hat{P}(\sqrt{\eta_2}\ket{1}_B-\sqrt{\eta_1(1-\eta_2)}\ket{2}_B),
\label{bobpovm2}
\\
\hat{\Pi}_{j=2,D_0}=&\hat{P}(\sqrt{(1-\eta_2)(1-\eta_1)}\ket{2}_B+\sqrt{\eta_2}\ket{3}_B),
\label{bobpovm3}
\\
\hat{\Pi}_{j=2,D_1}=&\hat{P}(\sqrt{\eta_2}\sqrt{1-\eta_1}\ket{2}_B-\sqrt{1-\eta_2}\ket{3}_B).
\label{bobpovm4}
\end{align}

\subsection{Security proof based on complementarity}
\label{bob:comple}
When a detection event occurs at time slot TS$j$, 
Alice applies the C-NOT gate on her systems $A_j$ and $A_{j+1}$ followed by 
measuring system $A_j$ in the $X$-basis to obtain her sifted key. 
In the security proof based on complementarity~\cite{koashi2009}, we are interested in how well Alice can predict the 
outcome $z_{A_j}\in\{0,1\}$ if system $A_j$ were measured in the $Z$-basis, 
complementary to the key generation basis ($X$-basis). 
We define the phase error event as those where Alice fails in predicting $z_{A_j}$. 
The ratio $e_{\ph}$ 
of the number of failure events $N_{\ph}$ to the number of code events $N_{\code}$ is called the phase error rate. 
Thanks to the following Theorem~\ref{th:keyrate} proven in~\cite{koashi2009}, 
the amount of bits to be shortened in the privacy amplification step is determined by $e_{\ph}$. 
\begin{theorem}
\label{th:keyrate}
If Alice and Bob shorten their reconciled keys of length $N_{\U{code}}$ by
\begin{align}
N_{\U{PA}}=N_{\code}h(e_{\ph})
\end{align}
in the privacy amplification step, they share a secret key of length
\begin{align}
\ell=N_{\code}-N_{\U{PA}}-N_{\U{EC}}.
\end{align}
Here, $h(x):=-x\log_2x-(1 - x)\log_2(1 - x)$, and 
$N_{\U{EC}}$ denotes the number of bits sacrificed in the bit error correction step. 
\end{theorem}
Given this theorem, our remaining task is to derive an upper bound on the phase error rate $e_{\ph}$
using experimentally observed data, such as the bit error rate $e_{\bit}$
\footnote{
\label{foot4}
Note that the role of the upper bound on the number of phase errors $N^U_{\rm ph}$
in the security proof with complementarity~\cite{koashi2009} 
is summarized as follows. 
Let $\bm{A}_{\rm sift}$ be Alice's qubits composed of $A_j$, which corresponds to $N_{\rm code}$ detection events. 
Alice obtains her secret key by applying the quantum circuit, which is constructed based on privacy amplification, 
followed by measuring $N_{\rm code}-N_{\rm PA}$ qubits in the $X$-basis. 
The goal of the virtual protocol is for Alice to make state $\ket{0}^{\otimes N_{\rm code}}$ (an eigenstate of 
the complementary observable) 
without changing the statistics of the final secret key. 
For this, we employ the random hashing idea. In this idea, the starting point is to note that 
when we have the upper bound on the number of phase errors (the number of wrong predictions of the 
complementary observable), 
the number of phase error patters if $\bm{A}_{\rm sift}$ 
were measured in the $Z$-basis is upper-bounded by $2^{N_{\rm code}h(N^U_{\rm ph}/N_{\rm code})}$. 
Hence, once Alice has the $N_{\rm PA}=N_{\rm code}h(N^U_{\rm ph}/N_{\rm code})$-bit syndrome information of the 
$Z$-basis measurement outcomes, she can uniquely identify the $Z$-basis measurement outcomes. 
This syndrome information is obtained by measuring $N_{\rm PA}$ qubits among $\bm{A}_{\rm sift}$ in the $Z$-basis 
after the quantum circuit based on privacy amplification, which are not measured to obtain the secret key. 
After Alice uniquely identifies the outcomes, she applies the Paul-$X$ operator (bit-flip operation)
to make all the qubits in $\ket{0}$, which achieves her goal. 
This is very important observation in converting the virtual protocol to the actual protocol. To see this, notice that Alice does not need the syndrome information because Alice can skip the correction step. At this point, all the measurements are in 
the $X$-basis, allowing Alice to directly measure all her qubits in the $X$-basis followed by classical data processing, i.e., 
privacy amplification, directed by the quantum circuit. This completes the conversion to the actual protocol.
}. 
The result can be stated as the following theorem.

\begin{theorem}
\label{theorem2}
In the asymptotic limit of large $N_{\det}$, 
the upper bound $e^U_{\ph}$ on the phase error rate $e_{\ph}=N_{\ph}/N_{\code}$ of our DPS protocol is given by
\begin{align}
e^U_{\ph}=\lambda(\eta_1^U,\eta_2^U)
\left(e_{\bit}+\frac{\sqrt{q_1q_3}}{Q}+2\delta^{(\U{BS})}_2\right)+\frac{q_2}{Q}
\label{def:theo2}
\end{align}
with
\begin{align}
\lambda(\eta_1,\eta_2):=
\frac{1-(1-\eta_1)\eta_2+\sqrt{
[1-(1-\eta_1)\eta_2]^2-4\eta_1(1-\eta_1)(1-\eta_2)^2}}
{2(1-\eta_1)(1-\eta_2)^2}.
\label{def:lambdade}
\end{align}
Note that $e_{\bit}$ and $Q$ are defined in Eq.~(\ref{eq:exdef}), while 
$q_n$ and $\delta^{(\U{BS})}_2$ are defined in Eqs.~(\ref{def:qn}) and (\ref{eq:etabound}), respectively. 
\end{theorem}
Combining Theorems~\ref{th:keyrate} and \ref{theorem2}, the secret key rate $R:=\ell/3N_\em$ 
per emitted pulse is expressed as
\begin{align}
R=\frac{N_{\code}[1-h(e^U_{\ph})]-N_{\U{EC}}}{3N_{\em}}.
\label{eq:keyr}
\end{align}
The rest of this section is devoted to proving Theorem~\ref{theorem2}. To achieve this, we begin by writing down the 
POVM elements for the occurrence of bit and phase errors.

\subsection{POVM elements for bit and phase error events}
\label{bob:povms}
\begin{figure*}[t]
\includegraphics[width=11cm]{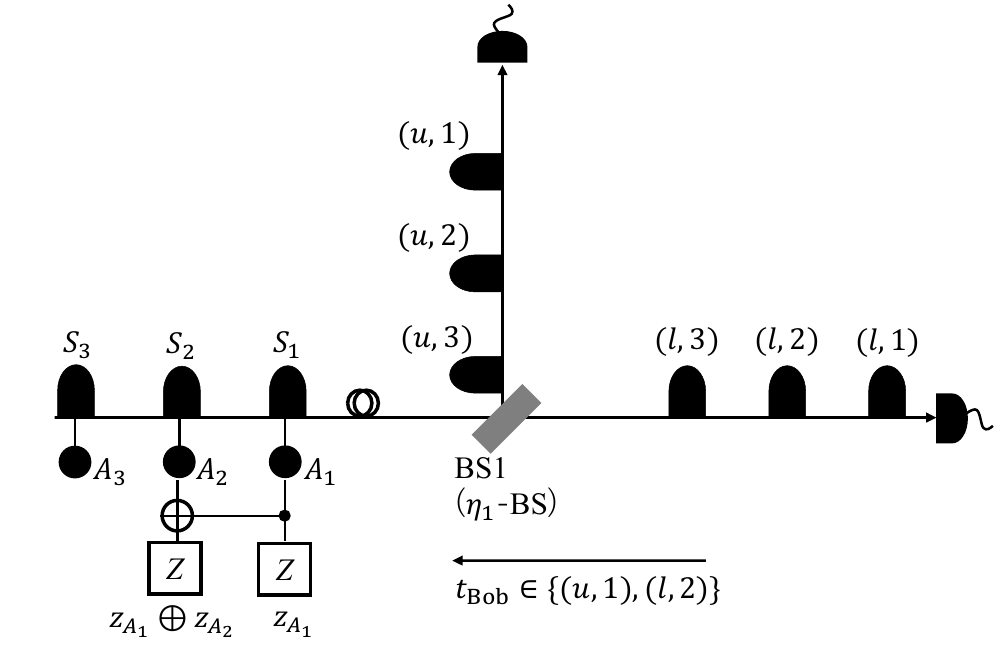}
\caption{
The setup to predict the complementary observable $z_{A_1}$ when a detection event occurs at TS1. 
Alice sends systems $S_1S_2S_3$ of state $\ket{\Phi}$ defined in Eq.~(\ref{fekj}) to Bob, and he splits the incoming pulses 
into two pulse trains by the BS1. To enhance the accuracy of Alice's prediction of the complementary observable, 
Bob measures which of the pulses has a single photon by removing the BS2 
and announces its result $t_{\U{Bob}}$ to Alice. 
When a detection event occurs at TS1, namely, $t_{\U{Bob}}\in\{(u,1),(l,2)\}$, 
Alice measures system $A_2$ in the $Z$-basis after applying the 
C-NOT gate and obtains the outcome $z_{A_1}\oplus z_{A_2}\in\{0,1\}$. 
Utilizing this parity information along with $t_{\U{Bob}}$, 
she predicts $z_{A_1}$, whose prediction strategy is described in Eqs.~(\ref{111}), (\ref{222}), and (\ref{333}). 
}
 \label{fig:virbob}
\end{figure*}

First, the POVM element associated with observing the bit error event at time slot TS$j$ is given by~\cite{dps2019}
\begin{align}
\hat{e}^{(\U{TS}j)}_{\bit}(\eta_1,\eta_2)=
&[\hat{P}(\hat{H}\ket{0}_{A_j}\hat{H}\ket{0}_{A_{j+1}})+\hat{P}(\hat{H}\ket{1}_{A_j}\hat{H}\ket{1}_{A_{j+1}})]
\otimes \hat{\Pi}_{j,D_1}\notag\\
+&
[\hat{P}(\hat{H}\ket{0}_{A_j}\hat{H}\ket{1}_{A_{j+1}})+\hat{P}(\hat{H}\ket{1}_{A_j}\hat{H}\ket{0}_{A_{j+1}})]
\otimes \hat{\Pi}_{j,D_0}.
\label{eq:laterebit}
\end{align}
By taking a sum over $j$, we obtain the POVM element for obtaining the bit error event as
\begin{align}
\label{koret1}
\hat{e}_{\bit}(\eta_1,\eta_2)=\hat{e}^{(\U{TS}1)}_{\bit}(\eta_1,\eta_2)+\hat{e}^{(\U{TS}2)}_{\bit}(\eta_1,\eta_2).
\end{align}

Next, we derive the POVM element for the occurrence of the phase error event, namely, the event of Alice 
failing to predict the complementary observable $z_{A_j}\in\{0,1\}$ 
(see Fig.~\ref{fig:virbob} for the setup to predict $z_{A_j}$). 
For this, we consider that Bob virtually removes the BS2 and measures which of the pulses has a single photon. 
This measurement is allowed because the classical information announced by Bob 
is the same as the one of the actual protocol
\footnote{This is because the POVM element for obtaining a detection event at TS1 [TS2] 
with the virtual measurement is $\ket{1}\bra{1}_B+\eta_1\ket{2}\bra{2}_B$ 
[$(1-\eta_1)\ket{2}\bra{2}_B+\ket{3}\bra{3}_B$], which is equivalent to $\hat{\Pi}_{1,D_0}+\hat{\Pi}_{1,D_1}$ 
[$\hat{\Pi}_{2,D_0}+\hat{\Pi}_{2,D_1}$] from Eqs.~(\ref{ertjklghertjgyh})-(\ref{bobpovm4}).}. 
In the following, 
we focus on the case where a detection event occurs at TS1 (with $j=1$), but the same discussion holds for TS2. 
For this, to enhance the accuracy of Alice's prediction of $z_{A_1}$, 
we consider that Bob announces the outcome of this virtual measurement, 
denoted by $t_{\rm Bob}\in\{(u,1),(l,2)\}$, and Alice utilizes this information to predict $z_{A_1}\in\{0,1\}$. 
Alice measures system $A_2$ as well in the $Z$-basis after applying the C-NOT gate and obtains the 
outcome $z_{A_1}\oplus z_{A_2}\in\{0,1\}$. 
Note that the detailed explanations of why Alice can utilize the information of $t_{\rm Bob}$ and $z_{A_1}\oplus z_{A_2}$ 
for predicting $z_{A_1}$ is found in Appendix~\ref{app:comp}. 
Once Alice has this parity information, she knows that $(z_{A_1},z_{A_2})$ belongs 
either to $\mathcal{I}_0:=\{(0,0), (1,1)\}$ or $\mathcal{I}_1:=\{(0,1), (1,0)\}$, depending on $z_{A_1}\oplus z_{A_2}$. 
As proven in Ref.~\cite{dps2019}, $z_{A_j}=1$ (0) approximately indicates that the $j$th emitted pulse had a single photon (zero photon)
\footnote{
This can be seen by rewriting the state in Eq.~(\ref{fekj}) using the $Z$-basis states as
\begin{align}
\sum_{b_j=0}^1
\frac{\hat{H}\ket{b_j}_{A_j}\ket{\psi_{b_j}}_{S_jR_j}}{\sqrt{2}}=
\frac{\ket{0}_{A_j}(\ket{\psi_0}_{S_jR_j}+\ket{\psi_1}_{S_jR_j})
+
\ket{1}_{A_j}(\ket{\psi_0}_{S_jR_j}-\ket{\psi_1}_{S_jR_j})
}{2}.
\end{align}
Since the probabilities of the emitted states being the vacuum state are assumed to be equal for both bits in 
Eq.~(\ref{eq:vace}), $\frac{\ket{\psi_0}_{S_jR_j}-\ket{\psi_1}_{S_jR_j}}{\sqrt{2[1-\U{Re}\expect{\psi_0|\psi_1}]}}$ 
contains at least one photon, while $\frac{\ket{\psi_0}_{S_jR_j}+\ket{\psi_1}_{S_jR_j}}{\sqrt{2[1+\U{Re}\expect{\psi_0|\psi_1}]}}$ contains zero or more photons. Utilizing the fact that the intensity of the emitted pulse is weak, the former state, corresponding to $z_{A_j}=1$, is approximately a single-photon state, while the latter state, corresponding to $z_{A_j}=0$, 
is approximately the vacuum state.
}. 
When $(z_{A_1},z_{A_2})\in\mathcal{I}_1$, it is reasonable to predict that the first or second emitted pulse had a single photon 
depending on $t_{\rm Bob}=(u,1)$ or $t_{\rm Bob}=(l,2)$, respectively. 
Hence, Alice's prediction of $z_{A_1}$ when $(z_{A_1},z_{A_2})\in\mathcal{I}_1$ is summarized as follows: 
\begin{align}
(z_{A_1},z_{A_2})=(1,0)~\U{if}~(z_{A_1},z_{A_2})\in\mathcal{I}_1~\wedge~&t_{\rm Bob}=(u,1),
\label{111}\\
(z_{A_1},z_{A_2})=(0,1)~\U{if}~(z_{A_1},z_{A_2})\in\mathcal{I}_1~\wedge~&t_{\rm Bob}=(l,2).
\label{222}
\end{align}
On the other hand, if $(z_{A_1},z_{A_2})\in\mathcal{I}_0$, 
as each emitted pulse is weak, the likelihood of both pulses emitting a single photon is lower than that of the two pulses being in the vacuum state. 
Therefore, it is reasonable to predict $z_{A_1}$ as zero independently of Bob's information of $t_B$. This prediction is described as
\begin{align}
 (z_{A_1},z_{A_2})=(0,0)~\U{if}~(z_{A_1},z_{A_2})\in\mathcal{I}_0~\wedge~&t_{\rm Bob}\in\{(u,1),(l,2)\}.
\label{333}
\end{align}
The phase error event at TS1 occurs when the prediction of $z_{A_1}$ fails in any of the cases described by 
Eqs.~(\ref{111}), (\ref{222}), or (\ref{333}), and hence the POVM element for obtaining the phase error event at TS1 can be expressed as
\begin{align}
\hat{e}_{\ph}^{(\U{TS1})}(\eta_1)=
&\hat{P}(\ket{0}_{A_1}\ket{1}_{A_2}\ket{1}_B)+\eta_1
\hat{P}(\ket{1}_{A_1}\ket{0}_{A_2}\ket{2}_B)
+\hat{P}(\ket{1}_{A_1}\ket{1}_{A_2})\otimes(\ket{1}\bra{1}_B+\eta_1\ket{2}\bra{2}_B)
\label{tyoi}\\
=&\hat{P}(\ket{1}_{A_2}\ket{1}_B)+\eta_1\hat{P}(\ket{1}_{A_1}\ket{2}_B).
\label{eq:ephpovmts1}
\end{align}
With the same discussion for TS2, we arrive at
\begin{align}
\hat{e}_{\ph}^{(\U{TS2})}(\eta_1)
=&(1-\eta_1)\hat{P}(\ket{1}_{A_3}\ket{2}_B)+\hat{P}(\ket{1}_{A_2}\ket{3}_B).
\label{eq:ephpovmts2}
\end{align}
Taking a sum of Eqs.~(\ref{eq:ephpovmts1}) and (\ref{eq:ephpovmts2}), the resulting 
POVM element for the phase error event is given by
\begin{align}
\hat{e}_{\ph}(\eta_1)=&\hat{e}_{\ph}^{(\U{TS1})}(\eta_1)+\hat{e}_{\ph}^{(\U{TS2})}(\eta_1)\notag\\
=&\sum_{\vec{a}\in\{0,1\}^3}\hat{P}(\ket{\vec{a}}_{\bm{A}})\otimes\left\{
\delta_{a_2,1}\ket{1}\bra{1}_B+
\left[\delta_{a_1,1}\eta_1+\delta_{a_3,1}(1-\eta_1)\right]\ket{2}\bra{2}_B+
\delta_{a_2,1}\ket{3}\bra{3}_B\right\},
\label{koret2}
\end{align}
where $\vec{a}:=a_1a_2a_3\in\{0,1\}^3$, and $\delta_{x,y}$ denotes the Kronecker delta. 

\subsection{Derivation of the number of phase errors}
\label{bob:derivation}
Here, we derive the upper bound $e^U_{\ph}$ on the phase error rate 
by establishing the relationship between 
POVM elements for the bit error and phase error events in Eqs.~(\ref{koret1}) and (\ref{koret2}). 
This derivation relies on the following two Lemmas~\ref{lemma1} and \ref{lemma2}, which are extensions of Lemmas 1 and 2 in~\cite{dps2019}, respectively. The proofs of these lemmas are provided in Appendices~\ref{app:l1} and~\ref{app:l2}.

\begin{lemma}
\label{lemma1}
For any $\eta_1\in \mathcal{R}_1=[\eta^{L}_1,\eta^{U}_1]$ and $\eta_2\in \mathcal{R}_2=[\eta^{L}_2,\eta^{U}_2]$, 
\begin{align}
\hat{P}_1\hat{e}_{\ph}(\eta_1)\hat{P}_1\le\lambda(\eta_1^U,\eta_2^U)
\left(\hat{P}_1\hat{e}_{\bit}(\eta_1,\eta_2)\hat{P}_1+2\delta^{(\U{BS})}_2\right)
\label{eq:lemma1}
\end{align}
holds. 
Here, $\hat{P}_1$ is a projector acting on system $\bm{A}:=A_1A_2A_3$, projecting onto the subspace 
with the weight of 1 along the $Z$-basis, namely, 
\begin{align}
\label{eq:defP1}
\hat{P}_1=\sum_{\vec{a}:\wt(\vec{a})=1}\hat{P}(\ket{\vec{a}}_{\bm{A}}).
\end{align}
\end{lemma}

\begin{lemma}
For any $\eta_1$ and $\eta_2$ of $0<\eta_1,\eta_2<1$ and any state $\hat{\sigma}$ of systems $A_1A_2A_3B$, 
we have
\begin{align}
\tr[\hat{P}_1\hat{e}_{\bit}(\eta_1,\eta_2)\hat{P}_1\hat{\sigma}]
\le\tr[\hat{e}_{\bit}(\eta_1,\eta_2)\hat{\sigma}]+\sqrt{\tr[\hat{P}_1\hat{\sigma}]\cdot\tr[\hat{P}_3\hat{\sigma}]},
\label{eq:lemma2}
\end{align}
where 
\begin{align}
\hat{P}_3:=\hat{P}(\ket{111}_{A_1A_2A_3}).
\label{eq:exp3}
\end{align}
\label{lemma2}
\end{lemma}
Using Lemmas~\ref{lemma1} and \ref{lemma2}, we calculate the upper bound on the probability of obtaining a phase error event. 
For this, we add $2\delta^{(\U{BS})}_2$ and then multiply by $\lambda(\eta^U_1,\eta^U_2)>0$ to Eq.~(\ref{eq:lemma2}), 
and we obtain 
\begin{align}
\tr\left[\lambda(\eta^U_1,\eta^U_2)\left(\hat{P}_1\hat{e}_{\bit}(\eta_1,\eta_2)\hat{P}_1+2\delta^{(\U{BS})}_2\right)\hat{\sigma}\right]
\le
\lambda(\eta^U_1,\eta^U_2)\left(
\tr[\hat{e}_{\bit}(\eta_1,\eta_2)\hat{\sigma}]
+\sqrt{\tr[\hat{P}_1\hat{\sigma}]\tr[\hat{P}_3\hat{\sigma}]}
+2\delta^{(\U{BS})}_2\right).
\end{align}
Applying Lemma~\ref{lemma1} to lower-bound the left-hand side results in
\begin{align}
\tr[\hat{P}_1\hat{e}_{\ph}(\eta_1)\hat{P}_1\hat{\sigma}]\le \lambda(\eta^U_1,\eta^U_2)\left(
\tr[\hat{e}_{\bit}(\eta_1,\eta_2)\hat{\sigma}]
+\sqrt{\tr[\hat{P}_1\hat{\sigma}]\tr[\hat{P}_3\hat{\sigma}]}
+2\delta^{(\U{BS})}_2\right).
\label{eq:rLemma1}
\end{align}
As can be seen from Eq.~(\ref{koret2}), Alice's operator in $\hat{e}_{\ph}(\eta_1)$ is diagonalized in the $Z$-basis, and 
$\hat{P}_0\hat{e}_{\ph}(\eta_1)\hat{P}_0=0$ holds for any $\eta_1$, where 
\begin{align}
\hat{P}_0:=\hat{P}(\ket{000}_{A_1A_2A_3}). 
\label{eq:exp0}
\end{align}
This implies
\begin{align}
\tr[\hat{e}_{\ph}(\eta_1)\hat{\sigma}]
=\sum_{a=0}^3\tr[\hat{P}_a\hat{e}_{\ph}(\eta_1)\hat{P}_a\hat{\sigma}]
\le
\tr[\hat{P}_1\hat{e}_{\ph}(\eta_1)\hat{P}_1\hat{\sigma}]+\tr[(\hat{P}_{2}+\hat{P}_3)\hat{\sigma}]
\label{eq:kkmdG}
\end{align}
with 
\begin{align}
\hat{P}_2:=\sum_{\vec{a}:\wt(\vec{a})=2}\hat{P}(\ket{\vec{a}}_{\bm{A}}). 
\label{eq:exp2} 
\end{align}
Applying Eq.~(\ref{eq:rLemma1}) to Eq.~(\ref{eq:kkmdG}) yields
\begin{align}
\tr[\hat{e}_{\ph}(\eta_1)\hat{\sigma}]
\le
\lambda(\eta^U_1,\eta^U_2)\left(
\tr[\hat{e}_{\bit}(\eta_1,\eta_2)\hat{\sigma}]
+\sqrt{\tr[\hat{P}_1\hat{\sigma}]\tr[\hat{P}_3\hat{\sigma}]}
+2\delta^{(\U{BS})}_2\right)+\tr[(\hat{P}_{2}+\hat{P}_3)\hat{\sigma}].
\label{eq:relationpros}
\end{align}
Once we obtain the upper bound on the probability of obtaining the phase error event, 
our remaining task is to transform it into the one in terms of corresponding random variables.

Let us consider that Alice and Bob sequentially measure their systems $\bm{A}=A_1A_2A_3$ and $B$ in order from the 
first detection event. 
For each $i$th trial $(1\le i\le N_{\det})$, the detection 
event is assigned to a code or sample event with probabilities $t$ and $1-t$, respectively. 
When the code event is chosen, Alice and Bob learn the weight $a$ with POVM $\{\hat{P}_a\}_{a=0}^3$ and 
whether they have a phase error or not with POVM $\{\hat{e}_{\ph}(\eta_1),\hat{I}_{\bm{A}B}-\hat{e}_{\ph}(\eta_1)\}$. 
The reason why these two measurements can be considered simultaneously is that these measurements commute, namely 
$[\hat{e}_{\ph}(\eta_1),\hat{P}_a]=0$, 
as can be seen from Eqs.~(\ref{koret2}), (\ref{eq:defP1}), (\ref{eq:exp3}), (\ref{eq:exp0}) and (\ref{eq:exp2}).

In the case of the sample event, Alice and Bob measure their systems with POVM $\{\hat{e}_{\bit}(\eta_1,\eta_2), 
\hat{I}_{\bm{A}B}-\hat{e}_{\bit}(\eta_1,\eta_2)\}$ to determine the presence of a bit error. 
The set $\Omega$ of the $i$th measurement outcome $\omega_i$ is then given by
\begin{align}
\Omega=\bigcup_{a=0}^3\{\ph\wedge a,\overline{\ph}\wedge a\}\cup\{\bit,\overline{\bit}\}. 
\label{eq:Omega}
\end{align}
Here, ``bit" $``(\overline{\bit})" $and ``ph" $``(\overline{\ph})"$ denote the outcomes of obtaining the bit error (no bit error)
and phase error (no phase error) events, respectively, and ``$a$" denotes the outcome of obtaining the weight $a$. 
According to $\omega_i$, we define the following three random variables:
\begin{align}
\chi^{(i)}_{\U{ph}}= 
\begin{cases} 1  & \mbox{if}~\omega_i\in\cup_{a=0}^3\{\U{ph}\wedge a\},\\
0 & \mbox{otherwise},
\end{cases}
\label{eq:cph}
\end{align}
\begin{align}
\chi^{(i)}_{a}= 
\begin{cases} 1  & \U{if}~\omega_i\in\{\U{ph}\wedge a,\overline{\U{ph}}\wedge a\},\\
0 & \mbox{otherwise},
\end{cases}
\label{eq:ca}
\end{align}
and
\begin{align}
\chi^{(i)}_{\U{bit}}= 
\begin{cases} 1  & \mbox{if}~\omega_i=\U{bit},\\
0 & \mbox{otherwise}.
\end{cases}
\label{eq:cbit}
\end{align}
We also define a non-decreasing sequence of $\sigma$ algebra $\{F^{(i)}\}_i$ with $i\in\{1,...,N_{\det}\}$ 
on sample space $\Omega^{\times N_{\det}}$ that identifies 
random variables including $\chi^{(j)}_{\U{ph}}$, $\{\chi^{(j)}_{a}\}_{a=0}^3$ and $\chi^{(j)}_{\U{bit}}$ for $j\in\{1,...,i\}$. 
Identifying one element of $F^{(i)}$ corresponds to identifying the first $i$ measurement outcomes, and hence the expectation 
$E[X^{(i)}|F^{(j)}]$ of the random variable $X^{(i)}$ conditional on $F^{(j)}$ is regarded as the expectation of $X^{(i)}$ conditioned on the first $j$ outcomes. 
The conditional expectations of $\chi^{(i)}_{\U{ph}}, \chi^{(i)}_{a}$ and $\chi^{(i)}_{\U{bit}}$ are respectively written as
 \begin{align}
E[\chi^{(i)}_{\U{ph}}|F^{(i-1)}]&=t\cdot\tr[\hat{e}_{\ph}(\eta_1)\hat{\sigma}^{F^{(i-1)}}_{\bm{A}B}],
\label{eq:ig1}
\\
E[\chi^{(i)}_{a}|F^{(i-1)}]&=t\cdot\tr[\hat{P}_{a}\hat{\sigma}^{F^{(i-1)}}_{\bm{A}B}], 
\label{eq:ig2}
\\
E[\chi^{(i)}_{\U{bit}}|F^{(i-1)}]&=(1-t)\cdot\tr[\hat{e}_{\bit}(\eta_1,\eta_2)\hat{\sigma}^{F^{(i-1)}}_{\bm{A}B}]. 
\label{eq:ig3}
\end{align}
Here, $\hat{\sigma}^{F^{(i-1)}}_{\bm{A}B}$ denotes the state of systems $\bm{A}B=A_1A_2A_3B$ conditioned on the 
first $i-1$ outcomes. 
Equation~(\ref{eq:relationpros}) can be rewritten in terms of the conditional expectations as 
\begin{align}
\frac{E[\chi^{(i)}_{\U{ph}}|F^{(i-1)}]}{t}\le
\lambda(\eta_1^U,\eta_2^U)\left(
\frac{E[\chi^{(i)}_{\U{bit}}|F^{(i-1)}]}{1-t}+\sqrt{\frac{E[\chi^{(i)}_1|F^{(i-1)}]}{t}\frac{E[\chi^{(i)}_3|F^{(i-1)}]}{t}}+2\delta^{(\U{BS})}_2
\right)+\sum_{a=2,3}\frac{E[\chi^{(i)}_a|F^{(i-1)}]}{t}.
\end{align}
Taking the sum from 1 to $N_{\det}$ and applying the Cauchy-Schwarz inequality give
\begin{align}
\label{eq:aftercw}
&\frac{\sum_{i=1}^{N_{\det}}E[\chi^{(i)}_{\U{ph}}|F^{(i-1)}]}{t}\le
\lambda(\eta_1^U,\eta_2^U)\left(
\frac{\sum_{i=1}^{N_{\det}}E[\chi^{(i)}_{\U{bit}}|F^{(i-1)}]}{1-t}+
\sqrt{\frac{\sum_{i=1}^{N_{\det}}E[\chi^{(i)}_1|F^{(i-1)}]}{t}\frac{\sum_{i=1}^{N_{\det}}E[\chi^{(i)}_3|F^{(i-1)}]}{t}}+2\delta^{(\U{BS})}_2N_{\det}\right)\notag\\
&+\sum_{i=1}^{N_{\det}}\sum_{a=2,3}\frac{E[\chi^{(i)}_a|F^{(i-1)}]}{t}.
\end{align}
To transform the sums of conditional expectations to the numbers of occurrences, we introduce the following random variables 
\begin{align}
X^{(i)}_{\xi}=&\sum_{j=1}^i\left(\chi^{(j)}_{\xi}-E[\chi^{(j)}_{\xi}|F^{(j-1)}]\right)
\label{eq:rX}
\end{align}
for $\xi\in\{\U{ph, bit},a\}$ and $i\in\{1,2,...,N_{\det}\}$. 
It is then straightforward to confirm that the sequence of random variables $\{X^{(i)}_{\xi}\}_{i=1}^{N_{\det}}$ 
is Martingale with the bounded difference condition, and hence Azuma's inequality~\cite{azumaineq} states that
\begin{align}
\U{Pr}[|X^{(N_{\det})}_{\xi}|>N_{\det}\zeta]\le2e^{-N_{\det}\zeta^2/2}
\end{align}
holds for any $N_{\det}>0$ and $\zeta>0$. 
In the asymptotic limit of large $N_{\det}$, we can disregard the deviation terms in Azuma's inequality. 
Consequently, each sum of the conditional expectations in Eq.~(\ref{eq:aftercw}) can be replaced with the respective 
random variables. In doing so, we obtain
\begin{align}
\frac{N_{\ph}}{tN_{\det}}\le\lambda(\eta_1^U,\eta_2^U)\left(\frac{N_{\bit}}{(1-t)N_{\det}}+
\frac{1}{N_{\det}}
\sqrt{\frac{N_{a=1}}{t}\frac{N_{a=3}}{t}}+2\delta^{(\U{BS})}_2\right)+\frac{N_{a\ge2}}{tN_{\det}}.
\label{eq:Nxi}
\end{align}
Here, we define $N_{\xi}:=\sum_{i=1}^{N_{\det}}\chi^{(i)}_\xi$ for $\xi\in\{\U{ph, bit},a\}$. 
As proven in Ref.~\cite{dps2019}, the upper bound on $N_{a\ge a'}$ is derived as $N_{a\ge a'}\le tN_{\em}q_{a'}$ 
in the asymptotic limit. 
Substituting these upper bounds results in
\begin{align}
\frac{N_{\ph}}{tN_{\det}}\le\lambda(\eta_1^U,\eta_2^U)\left(\frac{N_{\bit}}{(1-t)N_{\det}}+
\frac{N_{\em}}{N_{\det}}\sqrt{q_1q_3}+2\delta^{(\U{BS})}_2\right)+\frac{N_{\em}q_2}{N_{\det}}.
\label{eq:derapT}
\end{align}
In the asymptotic limit, as the number of code events $N_{\code}$ approaches $tN_{\det}$ 
and the number of sample events $N_{\sample}$ approaches $(1-t)N_{\det}$, we finally obtain 
the upper bound on the phase error rate as
\begin{align}
e^U_{\ph}=\lambda(\eta_1^U,\eta_2^U)\left(e_{\bit}+\frac{\sqrt{q_1q_3}}{Q}+2\delta^{(\U{BS})}_2\right)+\frac{q_2}{Q}.
\end{align}
This ends the proof of Theorem~\ref{theorem2}.

\section{simulation of key rate}
\label{sec:sim}
\begin{figure*}[t]
\includegraphics[width=9.5cm]{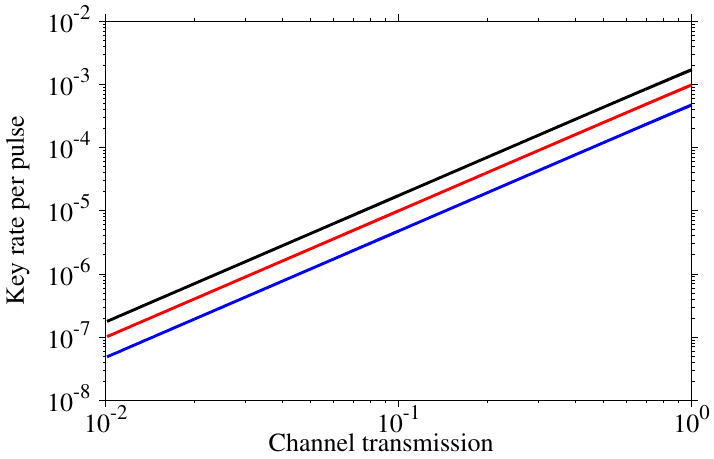}
\caption{
Secret key rate $R$ per emitted pulse as a function of the overall channel transmission $\eta$. 
The top, middle and bottom lines respectively represent the key rates under 
$\eta_1=\eta_2=0.5$, $\eta_1,\eta_2\in[0.5-0.005,0.5+0.005]$ and $\eta_1,\eta_2\in[0.5-0.01,0.5+0.01]$ 
with the bit error rate $e_{\bit}$ of 1\%. 
}
 \label{fig:kerate}
\end{figure*}
In this section, we present our simulation results of the key rate per emitted pulse given in Eq.~(\ref{eq:keyr}) 
as a function of the overall channel transmission $\eta$ including the detection efficiency. 
In the simulation, we assume that Alice employs a laser source emitting weak coherent pulses with the mean photon number 
$\mu$. 
In this case, $q_n$ defined in Eq.~(\ref{def:qn}) is written as $q_n=\sum_{m\ge n}e^{-3\mu}(3\mu)^m/m!$.
We suppose that the detection rate of the code events is $N_{\code}/N_{\em}=Q=2\eta\mu e^{-2\eta\mu}$, 
and the cost of bit error correction is $N_{\U{EC}}/N_{\em}=Qh(e_{\bit})$ with the bit error rate $e_{\bit}=1\%$. 
Regarding the transmittance of Bob's BS, 
we consider two situations; the one where the transmittance fluctuates within $\pm0.5\%$ of the ideal value 
(namely, $\mathcal{R}_1=\mathcal{R}_2=[0.5-0.005,0.5+0.005]$) 
and the other where the transmittance fluctuates within $\pm1\%$ (namely, 
$\mathcal{R}_1=\mathcal{R}_2=[0.5-0.01,0.5+0.01]$). 
In these cases, $\lambda(\eta_1^U,\eta_2^U)$ in Eq.~(\ref{def:theo2}) is $\lambda(0.505,0.505)\fallingdotseq5.41$ and 
$\lambda(0.51,0.51)\fallingdotseq5.60$, while $\lambda(0.5,0.5)\fallingdotseq5.24$ with the ideal transmittance. 
In Fig.~\ref{fig:kerate}, we optimize the key rate $R$ over $\mu$ for each transmission $\eta$. 
The top line represents the key rate assuming ideal BSs with the transmittance of 50\%~\cite{dps2019}. 
The middle (bottom) line shows the key rate under the fluctuation of $\pm$0.5\% ($\pm$1\%) in the transmittance of the BSs, 
resulting in a decrease to only 0.57 (0.27) times compared to the ideal case. 
These results clearly show that 
even under practical fluctuations in the transmittance of the BSs, the key rate of the DPS protocol 
does not degrade drastically, which strongly suggests the feasibility of the DPS protocol with practical measurement setups. 

\section{Conclusion}
\label{sec:conc}
In this paper, we have enhanced the implementation security of the differential-phase-shift (DPS) QKD protocol 
by providing a security proof incorporating a major imperfection in the measurement unit. 
Specifically, we take into account a practical imperfection in the beam splitters (BSs) inside Bob's Mach-Zehnder interferometer 
by only assuming that the transmittance surely lies within a certain range. 
Considering that it is feasible to manufacture the BSs with our assumption but not the one with exactly 50\% 
of the transmittance, our proof significantly relaxes the actual manufacturing process. 
As a result of our security proof, 
under a realistic assumption that the transmittance falls within a range of $\pm$0.5\% from the ideal value, 
we find that the secret key rate decreases to only 0.57 times lower than that with ideal beam splitters. 
Therefore, 
our result paves an important way to guarantee the implementation security of the DPS QKD with practical measurement setups.

We conclude this paper with some open questions.
\begin{enumerate}
\item
It is important to establish security proofs that simultaneously incorporate imperfections in the light source and measurement units. For instance, it is worth considering combining the results of this paper with the security proof of the DPS protocol using independent and identical light sources~\cite{dps2020}, or with the method of the security proof using correlated light sources~\cite{Pe2020}.
\item
Extending the current analysis to the finite-key analysis is an important work from a practical perspective. 
In so doing, we could apply the method of the finite-key analysis given in Ref.~\cite{dps2023}.
\item
It would be an interesting research topic to investigate whether our method of incorporating imperfections of the Mach-Zehnder interferometer into the security proof can be applied not only to the DPS protocol but also to other QKD protocols using Mach-Zehnder interferometers, such as DPS type protocols~\cite{chau,rrdps,hatake2017,dqps} 
and fully-passive QKD protocol~\cite{fpassive}. 
If other QKD protocols extract the secret key from detection events involving the receipt of a single photon, as is the case in this paper, then in their security proofs, it is crucial to construct POVM elements for bit and phase error events and to establish a relationship between these two elements. In this paper, these POVM elements were $\hat{e}_{\bit}(\eta_1,\eta_2)$ and $\hat{e}_{\ph}(\eta_1)$, as shown in Eqs.~(\ref{koret1}) and (\ref{koret2}), respectively, and the relationship between these two elements was derived in Eq.~(\ref{eq:relationpros}). 
When deriving the relationship between these two POVM elements for other QKD protocols, 
our analysis presented in Appendices 
in~\ref{app:l1} and \ref{app:l2}, which is the method of upper-bounding the probability of a phase error event using the probability of a bit error event under fluctuations in the transmittance of BSs, could be adopted.
\end{enumerate}

\section*{Acknowledgements}
We thank Go Kato and Yuki Takeuchi for helpful discussions. 
A.M. is partially supported by JST, ACT-X Grant No. JPMJAX210O, Japan 
and by JSPS KAKENHI Grant Number JP24K16977. 
K.T. acknowledges support from JSPS KAKENHI Grant Number 23H01096.

\appendix

\section{Proof of Theorem~\ref{theorem:bsposition}}
\label{app:pT}
\begin{figure*}[t]
\includegraphics[width=17cm]{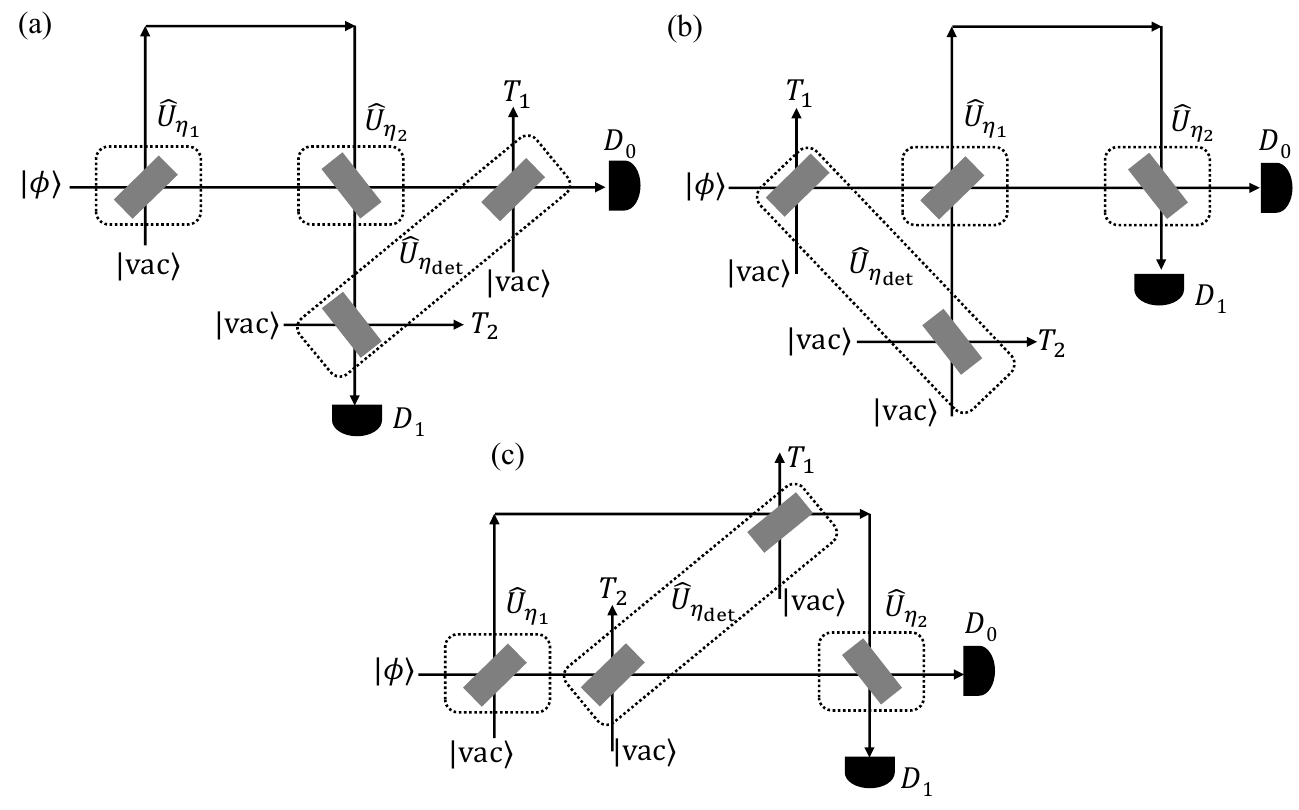}
\caption{
(a) Bob's actual measurement setup, where $\hat{U}_{\eta_{\det}}$ is placed just before the detectors. 
(b) Bob's measurement setup for proving the security of our DPS protocol, where $\hat{U}_{\eta_{\det}}$ is placed in front of the interferometer. 
(c) Bob's measurement setup used to prove Theorem~\ref{theorem:bsposition}, 
with $\hat{U}_{\eta_{\det}}$ placed in the interferometer. We prove the equivalence of (a) and (b) by showing that each is equivalent to (c).
}
 \label{fig:detection}
\end{figure*}

\begin{figure*}[t]
\includegraphics[width=13cm]{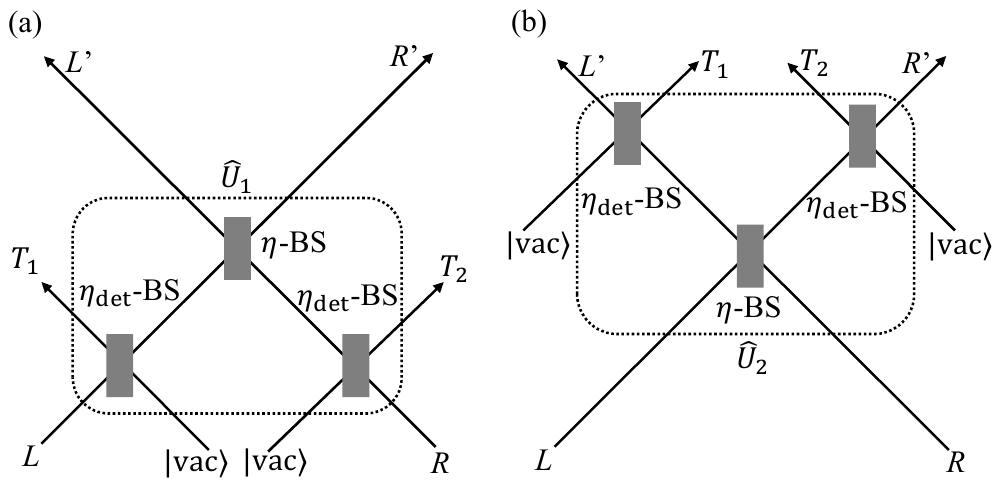}
\caption{
Time evolution of (a) $\hat{U}_1:=\hat{U}_{\eta}\hat{U}_{\eta_{\det}}$ and (b) $\hat{U}_2:=\hat{U}_{\eta_{\det}}\hat{U}_{\eta}$. 
The proof of Theorem~\ref{theorem:bsposition} 
is reduced to proving the equivalence of the states of modes $L'R'$ in both cases (a) and (b) 
for any input state $\ket{\psi}_{LR}$. 
}
\label{fig:detection2}
\end{figure*}
Here, we prove Theorem~\ref{theorem:bsposition}. 
We first recall that, as assumed in \ref{ass:b2}, the quantum efficiency $\eta_{\rm det}$ of the detectors can be modeled by a beam splitter with transmittance $\eta_{\det}$. 
Bob employs two PNR detectors, and the transformation of the two beam splitters is denoted by a single unitary operator 
$\hat{U}_{\det}$. 
We also define $\hat{U}_{\eta_1}$ and $\hat{U}_{\eta_2}$ as the unitary operators of the BSs with transmittance $\eta_1$ and $\eta_2$, respectively. 
The proof of Theorem~\ref{theorem:bsposition} is to show the following equation holds for any state $\ket{\phi}$:
\begin{align}
\tr_{T_1T_2}\hat{P}(\hat{U}_{\eta_{\det}}\hat{U}_{\eta_2}\hat{U}_{\eta_1}\ket{\phi}\ket{\vac})
=
\tr_{T_1T_2}\hat{P}(\hat{U}_{\eta_2}\hat{U}_{\eta_1}\hat{U}_{\eta_{\det}}\ket{\phi}\ket{\vac}).
\label{tgwg}
\end{align}
Here, $\ket{\phi}$ denotes the state of the one of the input modes, which is regarded as the input from Eve to Bob, 
while the state of the remaining three modes is in the vacuum state $\ket{\vac}$. 
Also, $T_1$ and $T_2$ represent the systems of the output modes of $\hat{U}_{\eta_{\det}}$. 
Figure~\ref{fig:detection} (a) and (b) respectively 
illustrate time evolution of the unitary operators $\hat{U}_{\eta_{\det}}\hat{U}_{\eta_2}\hat{U}_{\eta_1}$ and 
$\hat{U}_{\eta_2}\hat{U}_{\eta_1}\hat{U}_{\eta_{\det}}$ that appear in Eq.~(\ref{tgwg}). 
Equation~(\ref{tgwg}) indicates 
that the two states just before the detectors $D_0$ and $D_1$ under the setups of (a) and (b) are equivalent. 
To prove Eq.~(\ref{tgwg}), we introduce an intermediate setup, where $\hat{U}_{\eta_{\det}}$ is placed in the middle 
of $\hat{U}_{\eta_1}$ and $\hat{U}_{\eta_2}$. This setup is depicted in Fig.~\ref{fig:detection} (c), 
and we prove Eq.~(\ref{tgwg}) 
by showing that the setups of (a) and (b) are respectively equivalent to setup (c), namely, 
\begin{align}
\tr_{T_1T_2}\hat{P}(\hat{U}_{\eta_{\det}}\hat{U}_{\eta_2}\hat{U}_{\eta_1}\ket{\phi}\ket{\vac})
&=
\tr_{T_1T_2}\hat{P}(\hat{U}_{\eta_2}\hat{U}_{\eta_{\det}}\hat{U}_{\eta_1}\ket{\phi}\ket{\vac}),
\label{eq:secondbs}\\
\tr_{T_1T_2}\hat{P}(\hat{U}_{\eta_2}\hat{U}_{\eta_{\det}}\hat{U}_{\eta_1}\ket{\phi}\ket{\vac})
&=\tr_{T_1T_2}\hat{P}(\hat{U}_{\eta_2}\hat{U}_{\eta_1}\hat{U}_{\eta_{\det}}\ket{\phi}\ket{\vac})
.
\label{eq:firstbs}
\end{align}
By regarding $\hat{U}_{\eta_1}\ket{\phi}\ket{\vac}$ as the input states of 
$\hat{U}_{\eta_{\det}}\hat{U}_{\eta_2}$ and $\hat{U}_{\eta_2}\hat{U}_{\eta_{\det}}$ 
in Eq.~(\ref{eq:secondbs}) and by considering that Eq.~(\ref{eq:firstbs}) is equal to 
\begin{align}
\hat{U}_{\eta_2}\left(\tr_{T_1T_2}[\hat{P}(\hat{U}_{\eta_{\det}}\hat{U}_{\eta_{1}}\ket{\phi}\ket{\vac})]\right)
\hat{U}_{\eta_2}^\dagger=
\hat{U}_{\eta_2}
\left(\tr_{T_1T_2}[\hat{P}(\hat{U}_{\eta_1}\hat{U}_{\eta_{\det}}\ket{\phi}\ket{\vac})]\right)\hat{U}_{\eta_2}^\dagger,
\end{align}
we see that it is sufficient to prove the following equation for any $\ket{\psi}$:
\begin{align}
\tr_{T_1T_2}\hat{P}(\hat{U}_{\eta}\hat{U}_{\eta_{\det}}\ket{\psi}\ket{\vac})=
\tr_{T_1T_2}\hat{P}(\hat{U}_{\eta_{\det}}\hat{U}_{\eta}\ket{\psi}\ket{\vac}).
\label{yusyu}
\end{align}
Here, $\ket{\psi}=\sum_{n,m=0}^\infty x_{nm}\ket{n}\ket{m}$ with $x_{nm}\in\mathbb{C}$ 
represents any state of the two input modes among the four input modes. 
Once Eq.~(\ref{yusyu}) holds, Eqs.~(\ref{eq:secondbs}) and (\ref{eq:firstbs}) can be obtained by substituting 
$\ket{\psi}\ket{\vac}=\hat{U}_{\eta_1}\ket{\phi}\ket{\vac}$ and $\ket{\psi}\ket{\vac}=\ket{\phi}\ket{\vac}$, respectively. 
We call the case where $\eta_{\det}$-BSs are in front of the $\eta$-BS ``case (a)" and behind it ``case (b)". 
Figure~\ref{fig:detection2} illustrates time evolution of cases (a) and (b). 
Below, we separately calculate the states in these cases and show their equivalence. 
\\\\
{\bf Case (a) in Fig.~\ref{fig:detection2}: $\eta_{\det}$-BSs are in front of $\eta$-BS}\\
We define $\hat{U}_1:=\hat{U}_{\eta}\hat{U}_{\eta_{\det}}$ and calculate the left-hand side of Eq.~(\ref{yusyu}), i.e., 
$\tr_{T_1T_2}[\hat{P}(\hat{U}_1\ket{\psi}\ket{\vac})]$. $\hat{U}_1\ket{\psi}\ket{\vac}$ is calculated as
\begin{align}
\hat{U}_1\ket{\psi}\ket{\vac}
&=\sum_{n,m=0}^\infty x_{nm}\hat{U}_1\ket{n}\ket{m}\ket{\vac}\\
&=\sum_{n,m=0}^\infty\frac{x_{nm}}{\sqrt{n!m!}}(\hat{U}_1\hat{a}_L^\dagger \hat{U}_1^\dagger)^n(\hat{U}_1\hat{a}_R^\dagger \hat{U}_1^\dagger)^m\ket{\vac},
\label{uninini}
\end{align}
where $\hat{a}^\dagger_L$ and $\hat{a}^\dagger_R$ are the creation operators of input modes $L$ and $R$, respectively. 
We used $\hat{U}_1^{\dagger}\ket{\vac}=\ket{\vac}$ in the second equality, which holds 
because $\hat{U}_1$ is composed of BSs that never change the total number of photons. 
$\hat{a}_L^\dagger$ and $\hat{a}_R^\dagger$ evolve under $\hat{U}_1$ as
\begin{align}
\hat{U}_1\hat{a}_L^\dagger \hat{U}_1^\dagger&=
\sqrt{1-\eta_{\det}}\hat{a}_{T_1}^\dagger+
\sqrt{\eta_{\det}}\underbrace{
(\sqrt{1-\eta}\hat{a}_{L'}^\dagger+\sqrt{\eta}\hat{a}_{R'}^\dagger)
}_{=:\hat{A}},
\label{eq:i1}
\\
\hat{U}_1\hat{a}_R^\dagger \hat{U}_1^\dagger&=\sqrt{1-\eta_{\det}}\hat{a}_{T_2}^\dagger+
\sqrt{\eta_{\det}}\underbrace{
(\sqrt{\eta}\hat{a}_{L'}^\dagger-\sqrt{1-\eta}\hat{a}_{R'}^\dagger)
}_{=:\hat{A}'}.
\label{eq:i2}
\end{align}
Here, $L'$ and $R'$ denote the output modes of $\hat{U}_1$. 
Substituting Eqs.~(\ref{eq:i1}) and (\ref{eq:i2}) to Eq.~(\ref{uninini}) results in
\begin{align}
\hat{U}_1\ket{\psi}\ket{\vac}
&=\sum_{n,m=0}^\infty\frac{x_{nm}}{\sqrt{n!m!}}\left(
\sqrt{1-\eta_{\det}}\hat{a}_{T_1}^\dagger+\sqrt{\eta_{\det}}\hat{A}
\right)^n\left(
\sqrt{1-\eta_{\det}}\hat{a}_{T_2}^\dagger+\sqrt{\eta_{\det}}\hat{A}'\right)^m\ket{\vac}\\
&=\sum_{n,m=0}^\infty \frac{x_{nm}}{\sqrt{n!m!}}
\sum_{r=0}^n\binom{n}{r}(\sqrt{\eta_{\det}}\hat{A})^{n-r}
(\sqrt{1-\eta_{\det}}\hat{a}_{T_1}^\dagger)^r
\sum_{r'=0}^m\binom{m}{r'}(\sqrt{\eta_{\det}}\hat{A}')^{m-r'}
(\sqrt{1-\eta_{\det}}\hat{a}_{T_2}^\dagger)^{r'}\ket{\vac}\\
&=\sum_{n,m=0}^\infty \sum_{r=0}^n\sum_{r'=0}^m
\frac{x_{nm}}{\sqrt{n!m!}}
\binom{n}{r}\binom{m}{r'}(\sqrt{\eta_{\det}}\hat{A})^{n-r}(\sqrt{\eta_{\det}}\hat{A}')^{m-r'}
(\sqrt{1-\eta_{\det}}\hat{a}_{T_1}^\dagger)^r
(\sqrt{1-\eta_{\det}}\hat{a}_{T_2}^\dagger)^{r'}
\ket{\vac}.
\label{htej1}
\end{align}
By taking a partial trace over systems $T_1$ and $T_2$, we obtain 
the state in the left-hand side of Eq.~(\ref{yusyu}). 

Next, we calculate the right-hand side of Eq.~(\ref{yusyu}). \\\\
{\bf Case (b) in Fig.~\ref{fig:detection2}: $\eta_{\det}$-BSs are behind $\eta$-BS}\\
We define $\hat{U}_2:=\hat{U}_{\eta_{\det}}\hat{U}_{\eta}$ and calculate the right-hand side of Eq.~(\ref{yusyu}), i.e., 
$\tr_{T_1T_2}[\hat{P}(\hat{U}_2\ket{\psi}\ket{\vac})]$. 
Similar calculations to case (a) show that
\begin{align}
\hat{U}_2\ket{\psi}\ket{\vac}
&=\sum_{n,m=0}^\infty\frac{x_{nm}}{\sqrt{n!m!}}(\hat{U}_2\hat{a}_L^\dagger\hat{U}_2^\dagger)^n(\hat{U}_2\hat{a}_R^\dagger \hat{U}_2^\dagger)^m\ket{\vac}.
\label{khgakwhlr1}
\end{align}
The creation operators $\hat{a}_L^\dagger$ and $\hat{a}_R^\dagger$ evolve under $\hat{U}_2$ as
\begin{align}
\hat{U}_2\hat{a}_L^\dagger \hat{U}_2^\dagger&=
\sqrt{1-\eta_{\det}}(\sqrt{1-\eta}\hat{a}_{T_1}^\dagger+\sqrt{\eta}\hat{a}_{T_2}^\dagger)+
\sqrt{\eta_{\det}}\underbrace{
(\sqrt{\eta}\hat{a}_{R'}^\dagger+\sqrt{1-\eta}\hat{a}_{L'}^\dagger)}_{=\hat{A}},
\\
\hat{U}_2\hat{a}_R^\dagger \hat{U}_2^\dagger&=
\sqrt{1-\eta_{\det}}(\sqrt{\eta}\hat{a}_{T_1}^\dagger-\sqrt{1-\eta}\hat{a}_{T_2}^\dagger)+
\sqrt{\eta_{\det}}\underbrace{
(\sqrt{\eta}\hat{a}_{L'}^\dagger-\sqrt{1-\eta}\hat{a}_{R'}^\dagger)}_{=\hat{A}'}.
\end{align}
Substituting these equations to Eq. (\ref{khgakwhlr1}) results in
\begin{align}
&\hat{U}_2\ket{\psi}\ket{\vac}\notag\\
&=\sum_{n,m=0}^\infty\frac{x_{nm}}{\sqrt{n!m!}}\left[
\sqrt{\eta_{\det}}\hat{A}+\sqrt{1-\eta_{\det}}(\sqrt{1-\eta}\hat{a}_{T_1}^\dagger+\sqrt{\eta}\hat{a}_{T_2}^\dagger)
\right]^n\left[
\sqrt{\eta_{\det}}\hat{A}'+\sqrt{1-\eta_{\det}}(\sqrt{\eta}\hat{a}_{T_1}^\dagger-\sqrt{1-\eta}\hat{a}_{T_2}^\dagger)
\right]^m\ket{\vac}\\
&=\sum_{n,m=0}^\infty \frac{x_{nm}}{\sqrt{n!m!}}
\sum_{r=0}^n\binom{n}{r}(\sqrt{\eta_{\det}}\hat{A})^{n-r}
\left[\sqrt{1-\eta_{\det}}(\sqrt{1-\eta}\hat{a}_{T_1}^\dagger+\sqrt{\eta}\hat{a}_{T_2}^\dagger)\right]^r\notag\\
&\sum_{r'=0}^m\binom{m}{r'}(\sqrt{\eta_{\det}}\hat{A}')^{m-r'}
\left[\sqrt{1-\eta_{\det}}(\sqrt{\eta}\hat{a}_{T_1}^\dagger-\sqrt{1-\eta}\hat{a}_{T_2}^\dagger)\right]^{r'}
\ket{\vac}\\
&=\sum_{n,m=0}^\infty \sum_{r=0}^n\sum_{r'=0}^m\frac{x_{nm}}{\sqrt{n!m!}}
\binom{n}{r}\binom{m}{r'}(\sqrt{\eta_{\det}}\hat{A})^{n-r}(\sqrt{\eta_{\det}}\hat{A}')^{m-r'}\notag\\
&\left[\sqrt{1-\eta_{\det}}(\sqrt{1-\eta}\hat{a}_{T_1}^\dagger+\sqrt{\eta}\hat{a}_{T_2}^\dagger)\right]^r
\left[\sqrt{1-\eta_{\det}}(\sqrt{\eta}\hat{a}_{T_1}^\dagger-\sqrt{1-\eta}\hat{a}_{T_2}^\dagger)\right]^{r'}
\ket{\vac}.
\label{htej2}
\end{align}
By taking a partial trace over systems $T_1$ and $T_2$, 
we obtain $\tr_{T_1T_2}\hat{P}(\hat{U}_2\ket{\psi}\ket{\vac})$. 

Our aim is to prove Eq.~(\ref{yusyu}), i.e., $\tr_{T_1T_2}\hat{P}(\hat{U}_1\ket{\psi}\ket{\vac})=
\tr_{T_1T_2}\hat{P}(\hat{U}_2\ket{\psi}\ket{\vac})$. 
Since 
$\tr_{T_1T_2}\hat{P}(\hat{U}_1\ket{\psi}\ket{\vac})=
\tr_{T_1T_2}\hat{P}(\hat{W}_{T_1T_2}\hat{U}_1\ket{\psi}\ket{\vac})$ holds for any unitary operator 
$\hat{W}_{T_1T_2}$ acting on systems 
$T_1$ and $T_2$, by setting $\hat{W}_{T_1T_2}$ as the unitary operator of the BS with transmittance $\eta$, we have 
$\hat{W}_{T_1T_2}\hat{U}_1\ket{\psi}\ket{\vac}=\hat{U}_2\ket{\psi}\ket{\vac}$. 
This ends the proof of Theorem~\ref{theorem:bsposition}.

\section{Derivations of Eqs.~(\ref{ertjklghertjgyh})-(\ref{bobpovm4})}
\label{app:devMP}
In this section, we derive POVM $\{\hat{\Pi}_{j,D_i}\}_{j=1,2,i=0,1}$ in 
Eqs.~(\ref{ertjklghertjgyh})-(\ref{bobpovm4}). Let $l$ and $u$ ($l'$ and $u'$) denote the two input (output) modes of the BS2. 
See Fig.~\ref{fig:actual} for the setup. 
The BS2 evolves the annihilation operators $\hat{a}_{l}$ and $\hat{a}_{u}$ of the two input modes as follows:
\begin{align}
\hat{a}_{l}&\rightarrow\sqrt{\eta_2}\hat{a}_{l'}-\sqrt{1-\eta_2}\hat{a}_{u'},
\label{U1}
\\
\hat{a}_{u}&\rightarrow\sqrt{\eta_2}\hat{a}_{u'}+\sqrt{1-\eta_2}\hat{a}_{l'}.
\end{align}
The inverse of this transformation is given by
\begin{align}
\hat{a}_{l'}\rightarrow\sqrt{\eta_2}\hat{a}_{l}+\sqrt{1-\eta_2}\hat{a}_{u},
\label{usingKAO}
\\
\hat{a}_{u'}\rightarrow\sqrt{\eta_2}\hat{a}_{u}-\sqrt{1-\eta_2}\hat{a}_{l}.
\label{usingKAO2}
\end{align}
Using Eqs.~(\ref{usingKAO}) and (\ref{usingKAO2}), we construct $\hat{\Pi}_{j,D_i}$ for $i\in\{0,1\}$ and $j\in\{1,2\}$, which 
represents a POVM element for observing one photon in detector $D_i$ at time slot TS$j$. 
This operator corresponds to the projector 
$\hat{P}(\hat{a}^{\dagger}_{\U{TS}j,D_i}\ket{\vac})$ with $\hat{a}^{\dagger}_{\U{TS}j,D_0}$  
($\hat{a}^{\dagger}_{\U{TS}j,D_1}$) denoting the creation operator of mode $l'$ (mode $u'$) at TS$j$. 
Below, we express this projector in the basis $\mathcal{B}:=\{\ket{1}_B,\ket{2}_B,\ket{3}_B\}$. 

By the inverse transformation of the BS2, single photon states $\hat{a}^{\dagger}_{\U{TS}j,D_i}\ket{\U{vac}}$ evolve 
as follows: 
\begin{align}
\hat{a}^{\dagger}_{\U{TS1},D_0}\ket{\vac}
&\rightarrow\left(\sqrt{\eta_2}\hat{a}^{\dagger}_{l,2}+\sqrt{1-\eta_2}\hat{a}^{\dagger}_{u,1}\right)\ket{\vac}=
\sqrt{\eta_2}\hat{a}^{\dagger}_{l,2}\ket{\vac}+\sqrt{1-\eta_2}\ket{1},\\
\hat{a}^{\dagger}_{\U{TS1},D_1}\ket{\vac}
&\rightarrow
\left(\sqrt{\eta_2}\hat{a}^\dagger_{u,1}-\sqrt{1-\eta_2}\hat{a}^\dagger_{l,2}\right)\ket{\vac}
=\sqrt{\eta_2}\ket{1}-\sqrt{1-\eta_2}\hat{a}^\dagger_{l,2}\ket{\vac},\\
\hat{a}^{\dagger}_{\U{TS2},D_0}\ket{\vac}
&\rightarrow
\left(\sqrt{\eta_2}\hat{a}^\dagger_{l,3}+\sqrt{1-\eta_2}\hat{a}^\dagger_{u,2}\right)\ket{\vac}
=\sqrt{\eta_2}\ket{3}+\sqrt{1-\eta_2}\hat{a}^\dagger_{u,2}\ket{\vac},\\
\hat{a}^{\dagger}_{\U{TS2},D_1}\ket{\vac}
&
\rightarrow
\left(\sqrt{\eta_2}\hat{a}^\dagger_{u,2}-\sqrt{1-\eta_2}\hat{a}^\dagger_{l,3}
\right)\ket{\vac}
=\sqrt{\eta_2}\hat{a}^\dagger_{u,2}\ket{\vac}-\sqrt{1-\eta_2}\ket{3}.
\end{align}
Here, $\hat{a}^{\dagger}_{u,n}$ for $n=1,2$ 
[$\hat{a}^{\dagger}_{l,n}$ for $n=2,3$] represents 
the creation operator of the pulse $(u,n)$ [pulse $(l,n)$] that is the 
$n$th pulse passing through the upper arm (lower arm) of the Mach-Zehnder interferometer. 
The first and third equations follow from Eq.~(\ref{usingKAO}), while the second and fourth equations follow from 
Eq.~(\ref{usingKAO2}). 
Also, by the inverse transformation of the BS1, $\hat{a}^{\dagger}_{l,2}\ket{\vac}$ and 
$\hat{a}^{\dagger}_{u,2}\ket{\vac}$ respectively change to 
$\sqrt{\eta_1}\hat{a}^\dagger_{2}\ket{\vac}=\sqrt{\eta_1}\ket{2}_B$ 
and $\sqrt{1-\eta_1}\hat{a}^\dagger_{2}\ket{\vac}=\sqrt{1-\eta_1}\ket{2}_B$ 
with $\hat{a}^\dagger_{2}$ being the creation operator of the second pulse input to the BS1. 
Note that the term associated with the other input mode to the BS1 is ignored here. 
This is allowed because the state of that mode is always the vacuum, which is orthogonal to a single-photon state, and as a result the term has no effect on the measurement statistics in the DPS experiment. 
As a result, $\hat{\Pi}_{j,D_i}$ are expressed in the basis $\mathcal{B}:=\{\ket{1}_B,\ket{2}_B,\ket{3}_B\}$ 
as shown in Eqs.~(\ref{ertjklghertjgyh})-(\ref{bobpovm4}).

\section{Two hints for predicting Alice's complementarity observable}
\label{app:comp}
In this Appendix, we explain why Alice can use the information of the two outcomes obtained by the following 
measurements \ref{o1} and \ref{o2} to predict the complementary observable $z_{A_j}\in\{0,1\}$.
\begin{enumerate}[label=M\arabic*]
\item 
\label{o1}
Bob's measurement to determine which pulse contains a single photon between the pulses $(u,1)$ and 
$(l,2)$ when a detection event occurs at TS1. 
Similarly, between the pulses $(u,2)$ and $(l,3)$ for TS2.
\item
\label{o2}
Alice's measurement of the parity information $z_{A_j}\oplus z_{A_{j+1}}\in\{0,1\}$.
\end{enumerate}
For explanation purpose, let $\bm{A}_{\rm sift}$ denote Alice's qubits composed of $A_j$, which corresponds to 
$N_{\rm code}$ detection events, and $\hat{\rho}_{\bm{A}_{\rm sift}E}$ denote the state of $\bm{A}_{\rm sift}$ and 
Eve's system just before executing privacy amplification. 
The estimation of the number of phase errors is performed on this state by considering the $Z$-basis measurement on 
$\bm{A}_{\rm sift}$, a measurement complementary to the $X$-basis measurement. 
The point here is that even if Bob performs the virtual measurement \ref{o1} instead of the actual one, Bob 
can publicly announce the same information as the actual protocol (in particular, the information of 
which time slot (TS1 or TS2) he obtains the detection event), and hence the resulting state 
$\hat{\rho}_{\bm{A}_{\rm sift}E}$ remains the same. 

Bob's measurement \ref{o1} can be executed by measuring his system $B$, and 
Alice's measurement \ref{o2} can be done by measuring qubit $A_{j+1}$. 
Since measurements on different systems commute, 
these measurements commute with the measurement on qubit $A_j$ to learn the presence of a phase error. 
Therefore, we can define three random variables representing the outcomes of these measurements simultaneously, 
and when estimating the number of phase errors to this state $\hat{\rho}_{\bm{A}_{\rm sift}E}$, 
Alice can classify the occurrence of the phase error event according to each value of these random variables. 
This is the reason why we can introduce \ref{o1} and \ref{o2} for predicting $z_{A_j}$ and 
also the reason why $z_{A_j}$ is predicted according to $\mathcal{I}_0$ or $\mathcal{I}_1$ and according to the value of 
$t_{\rm Bob}$ in Eqs.~(\ref{111})-(\ref{333}).

\section{Proof of Lemma~\ref{lemma1}}
\label{app:l1}
In this section, we prove Lemma~\ref{lemma1}. We consider upper-bounding 
\begin{align}
\hat{P}_1(\hat{e}_{\ph}(\eta_1)-\lambda\hat{e}_{\bit}(\eta_1,\eta_2))\hat{P}_1
\label{P1ephP1}
\end{align}
with $\lambda>0$, where $\hat{e}_{\bit}(\eta_1,\eta_2)$, $\hat{e}_{\ph}(\eta_1)$ and $\hat{P}_1$ are defined in 
Eqs.~(\ref{koret1}), (\ref{koret2}) and (\ref{eq:defP1}), respectively. 
For this, it is convenient to introduce the following unitary operator
\begin{align}
\hat{U}_{AB}:=\sum_{i=1}^3 \hat{X}_{A_i}\otimes\ket{i}\bra{i}_B
\label{eq:defUAB}
\end{align}
with $\hat{X}_{A_i}:=\ket{+}\bra{+}_{A_i}-\ket{-}\bra{-}_{A_i}$ and evaluate the upper bound on
\begin{align}
(\hat{U}_{AB}\hat{P}_1\hat{U}_{AB}^\dagger)(\hat{U}_{AB}\hat{e}_{\ph}(\eta_1)\hat{U}_{AB}^\dagger-\lambda
\hat{U}_{AB}\hat{e}_{\bit}(\eta_1,\eta_2)\hat{U}_{AB}^\dagger)(\hat{U}_{AB}\hat{P}_1\hat{U}_{AB}^\dagger).
\label{eq:UP1U}
\end{align}
This operator $\hat{U}_{AB}$ is also introduced to prove the security of the DPS protocol with blockwise phase randomization~\cite{dps2012,dps2017}. 
From Eq.~(\ref{eq:defUAB}), it is straightforward to see that the following equations hold. 
\begin{align}
\hat{U}_{AB}\bigotimes_{j=1}^3\hat{H}\ket{s_j}_{A_j}\ket{i}_B
=(-1)^{s_i}\bigotimes_{j=1}^3\hat{H}\ket{s_j}_{A_j}\ket{i}_B
\label{defUn}
\end{align}
with $s_j\in\{0,1\}$ and 
\begin{align}
\hat{U}_{AB}\ket{s_1}_{A_1}\ket{s_2}_{A_2}\ket{s_3}_{A_3}\ket{i}_B=
\ket{s_i\oplus 1}_{A_i}\ket{s_j}_{A_j}\ket{s_k}_{A_k}\ket{i}_B
\label{eq:UephU}
\end{align}
with $i\in\{1,2,3\}$ and $j\neq k\in\{1,2,3\}\setminus\{i\}$. 

First, we calculate $\hat{U}_{AB}\hat{e}_{\bit}(\eta_1,\eta_2)\hat{U}_{AB}^\dagger$. 
Using Eqs.~(\ref{eq:laterebit}) and (\ref{defUn}) gives 
\begin{align}
\hat{U}_{AB}\hat{e}^{(\U{TS1})}_{\bit}(\eta_1,\eta_2)\hat{U}_{AB}^\dagger=&
\left[\hat{P}(\hat{H}\ket{0}_{A_1}\hat{H}\ket{0}_{A_2})+
\hat{P}(\hat{H}\ket{1}_{A_1}\hat{H}\ket{1}_{A_2})\right]\otimes\hat{\Pi}_{1,D_1}\notag\\
+&\left[\hat{P}(\hat{H}\ket{0}_{A_1}\hat{H}\ket{1}_{A_2})+
\hat{P}(\hat{H}\ket{1}_{A_1}\hat{H}\ket{0}_{A_2})\right]\otimes\hat{\Pi}^{\U{minus}}_{1,D_0}
\label{UebitU1}
\end{align}
with
\begin{align}
\hat{\Pi}^{\U{minus}}_{1,D_0}:=\hat{P}(\sqrt{1-\eta_2}\ket{1}_B-\sqrt{\eta_1\eta_2}\ket{2}_B). 
\end{align}
Note that $\hat{\Pi}^{\U{minus}}_{1,D_0}$ is equal to $\hat{\Pi}_{1,D_1}$ if $\eta_1=\eta_2=0.5$, 
and $\hat{\Pi}^{\U{minus}}_{1,D_0}\neq\hat{\Pi}_{1,D_1}$ otherwise. 
By defining the projector $\hat{Q}^{A_iA_{i+1}}_{kk}:=\hat{P}(\hat{H}\ket{k}_{A_i}\hat{H}\ket{k}_{A_{i+1}})$ for $k\in\{0,1\}$, 
Eq.~(\ref{UebitU1}) is rewritten as 
\begin{align}
\hat{U}_{AB}\hat{e}^{(\U{TS1})}_{\bit}(\eta_1,\eta_2)\hat{U}_{AB}^\dagger
&=\hat{I}_{\bm{A}}\otimes \hat{\Pi}^{\U{minus}}_{1,D_0}+(\hat{Q}^{A_1A_2}_{00}+\hat{Q}^{A_1A_2}_{11})\otimes (\hat{\Pi}_{1,D_1}-\hat{\Pi}^{\U{minus}}_{1,D_0})\\
&=\hat{I}_{\bm{A}}\otimes \hat{\Pi}^{\U{minus}}_{1,D_0}+(\hat{Q}^{A_1A_2}_{00}+\hat{Q}^{A_1A_2}_{11})\otimes 
\underbrace{\left[(2\eta_2-1)\ket{1}\bra{1}_B+\eta_1(1-2\eta_2)\ket{2}\bra{2}_B\right]}_{=:\hat{M}_{12}}.
\end{align}
By performing the analogous calculation for the other time slot TS2, we obtain
\begin{align}
\hat{U}_{AB}\hat{e}^{(\U{TS2})}_{\bit}(\eta_1,\eta_2)\hat{U}_{AB}^\dagger
=\hat{I}_{\bm{A}}\otimes \hat{\Pi}^{\U{minus}}_{2,D_0}+(\hat{Q}^{A_2A_3}_{00}+\hat{Q}^{A_2A_3}_{11})\otimes
\underbrace{\left[(1-\eta_1)(2\eta_2-1)\ket{2}\bra{2}_B+(1-2\eta_2)\ket{3}\bra{3}_B\right]}_{=:\hat{M}_{23}}
\end{align}
with
\begin{align}
\hat{\Pi}^{\U{minus}}_{2,D_0}:=\hat{P}(\sqrt{(1-\eta_2)(1-\eta_1)}\ket{2}_B-\sqrt{\eta_2}\ket{3}_B).
\end{align}
Taking the sum of $\hat{U}_{AB}\hat{e}^{(\U{TS1})}_{\bit}(\eta_1,\eta_2)\hat{U}_{AB}^\dagger$ and 
$\hat{U}_{AB}\hat{e}^{(\U{TS2})}_{\bit}(\eta_1,\eta_2)\hat{U}_{AB}^\dagger$ yields
\begin{align}
\hat{U}_{AB}\hat{e}_{\bit}(\eta_1,\eta_2)\hat{U}_{AB}^\dagger
=\hat{I}_{\bm{A}}\otimes \hat{\Pi}^{\U{minus}}+
(\hat{Q}^{A_1A_2}_{00}+Q^{A_1A_2}_{11})\otimes \hat{M}_{12}+
(\hat{Q}^{A_2A_3}_{00}+Q^{A_2A_3}_{11})\otimes \hat{M}_{23}
\label{eq:ebitU}
\end{align}
with
\begin{align}
\hat{\Pi}^{\U{minus}}:=\hat{\Pi}^{\U{minus}}_{1,D_0}+\hat{\Pi}^{\U{minus}}_{2,D_0}=
\begin{pmatrix}
1-\eta_2 & -\sqrt{\eta_1\eta_2(1-\eta_2)} &0\\
-\sqrt{\eta_1\eta_2(1-\eta_2)} &\eta_1\eta_2+(1-\eta_1)(1-\eta_2) &-\sqrt{\eta_2(1-\eta_1)(1-\eta_2)}\\
0&-\sqrt{\eta_2(1-\eta_1)(1-\eta_2)}& \eta_2
\end{pmatrix}\ge0.
\label{eq:MatrixPi}
\end{align}
Here, this matrix is represented in the basis $\mathcal{B}=\{\ket{1}_B,\ket{2}_B,\ket{3}_B\}$. 

By applying $\hat{M}_{12}\ge-|2\eta_2-1|\left(\ket{1}\bra{1}_B+\eta_1\ket{2}\bra{2}_B\right)$, 
$\hat{M}_{23}\ge-|2\eta_2-1|\left[(1-\eta_1)\ket{2}\bra{2}_B+\ket{3}\bra{3}_B\right]$ and 
$\hat{Q}^{A_iA_{i+1}}_{00}+\hat{Q}^{A_iA_{i+1}}_{11}\le \hat{I}_{\bm{A}}$ to Eq.~(\ref{eq:ebitU}), 
we have that $\hat{U}_{AB}\hat{e}_{\bit}(\eta_1,\eta_2)\hat{U}_{AB}^\dagger$ is lower-bounded as 
\begin{align}
\hat{U}_{AB}\hat{e}_{\bit}(\eta_1,\eta_2)\hat{U}_{AB}^\dagger
\ge\hat{I}_{\bm{A}}\otimes\hat{\Pi}^{\U{minus}}-|2\eta_2-1|.
\label{ayasuki}
\end{align}

Next, we calculate $\hat{U}_{AB}\hat{e}_{\ph}(\eta_1)\hat{U}_{AB}^\dagger$. For this, $\hat{e}_{\ph}(\eta_1)$ in Eq.~(\ref{koret2}) is 
expressed as
\begin{align}
\hat{e}_{\ph}(\eta_1)=\sum_{\vec{a}\in\{0,1\}^3}\hat{P}(\ket{\vec{a}}_{\bm{A}})\otimes\hat{\Pi}^{\ph}_{\vec{a}}
\label{eq:ephre}
\end{align}
with 
\begin{align}
\hat{\Pi}^{\ph}_{\vec{a}}:=\delta_{a_2,1}\ket{1}\bra{1}_B+[\delta_{a_1,1}\eta_1+\delta_{a_3,1}(1-\eta_1)]\ket{2}\bra{2}_B
+\delta_{a_2,1}\ket{3}\bra{3}_B\ge0.
\label{def:piph}
\end{align}
It is straightforward to confirm from Eq.~(\ref{eq:UephU}) that $\hat{e}_{\ph}(\eta_1)$ is invariant under $\hat{U}_{AB}$, i.e.,
\begin{align}
\hat{U}_{AB}\hat{e}_{\ph}(\eta_1)\hat{U}_{AB}^\dagger=\hat{e}_{\ph}(\eta_1).
\label{aergarw}
\end{align}
Also, direct calculation employing Eqs.~(\ref{eq:defP1}) and (\ref{eq:UephU}) shows that
\begin{align}
\hat{U}_{AB}\hat{P}_1\hat{U}_{AB}^\dagger
=\hat{P}(\ket{000}_{\bm{A}})\otimes \hat{I}_B+\sum_{\vec{a}:\U{wt}(\vec{a})=2}\hat{P}(\ket{\vec{a}}_{\bm{A}})\otimes 
\underbrace{\sum_{i=1}^3\hat{P}(\ket{i}_B)\delta_{a_i,1}}_{=:\hat{P}_{\vec{a}}}.
\label{sfdkj}
\end{align}
By substituting Eqs.~(\ref{ayasuki}), (\ref{aergarw}) and (\ref{sfdkj}) to Eq.~(\ref{eq:UP1U}), the upper bound 
on Eq.~(\ref{eq:UP1U}) is calculated as follows:
\begin{align}
&
\left(\hat{P}(\ket{000}_{\bm{A}})\otimes \hat{I}_B+\sum_{\vec{a}:\wt(\vec{a})=2}
\hat{P}(\ket{\vec{a}}_{\bm{A}})\otimes \hat{P}_{\vec{a}}\right)
\sum_{\vec{a}\in\{0,1\}^3}\hat{P}(\ket{\vec{a}}_{\bm{A}})\otimes
\left[\hat{\Pi}^{\ph}_{\vec{a}}-\lambda(\hat{\Pi}^{\U{minus}}-|2\eta_2-1|)
\right]
\notag\\
&\left(\hat{P}(\ket{000}_{\bm{A}})\otimes \hat{I}_B+\sum_{\vec{a}:\wt(\vec{a})=2}\hat{P}(\ket{\vec{a}}_{\bm{A}})\otimes \hat{P}_{\vec{a}}\right)\\
=&
\hat{P}(\ket{000}_{\bm{A}})\otimes\left[
\hat{\Pi}^{\ph}_{000}-\lambda(\hat{\Pi}^{\U{minus}}-|2\eta_2-1|)\right]+
\sum_{\vec{a}:\wt(\vec{a})=2}\hat{P}(\ket{\vec{a}}_{\bm{A}})\otimes
\hat{P}_{\vec{a}}\left[\hat{\Pi}^{\ph}_{\vec{a}}-\lambda(\hat{\Pi}^{\U{minus}}-|2\eta_2-1|)\right]\hat{P}_{\vec{a}}\\
\le&
\lambda|2\eta_2-1|\left(\hat{P}(\ket{000}_{\bm{A}})
\otimes\hat{I}_B+\sum_{\vec{a}:\wt(\vec{a})=2}\hat{P}(\ket{\vec{a}}_{\bm{A}})\otimes
\hat{P}_{\vec{a}}
\right)+
\sum_{\vec{a}:\wt(\vec{a})=2}\hat{P}(\ket{\vec{a}}_{\bm{A}})\otimes\hat{T}_{\vec{a}}
\label{eq:ph000}
\\
\le&
\lambda|2\eta_2-1|+
\sum_{\vec{a}:\wt(\vec{a})=2}\hat{P}(\ket{\vec{a}}_{\bm{A}})\otimes\hat{T}_{\vec{a}}.
\label{kyoaya}
\end{align}
The first inequality follows by $\hat{\Pi}^{\ph}_{000}=0$ from Eq.~(\ref{def:piph}), 
$-\lambda\hat{P}(\ket{000}_{\bm{A}})\otimes\hat{\Pi}^{\U{minus}}\le0$ and the definition 
\begin{align}
\hat{T}_{\vec{a}}:=\hat{P}_{\vec{a}}\left[\hat{\Pi}^{\ph}_{\vec{a}}-\lambda\hat{\Pi}^{\U{minus}}\right]\hat{P}_{\vec{a}}. 
\end{align}
The second inequality follows because 
$\hat{P}(\ket{000}_{\bm{A}})\otimes\hat{I}_B+\sum_{\vec{a}:\wt(\vec{a})=2}\hat{P}(\ket{\vec{a}}_{\bm{A}})\otimes
\hat{P}_{\vec{a}}\le \hat{I}_{\bm{A}B}$. 

Below, we evaluate the largest eigenvalue of $\hat{T}_{\vec{a}}$ for $\vec{a}=110,101$ and 011 of $\wt(\vec{a})=2$. 
\\
(i) First, we consider the case of  $\vec{a}=110$. $\hat{T}_{110}$ is expressed as
\begin{align}
\hat{T}_{110}
=
[1-\lambda(1-\eta_2)]\ket{1}\bra{1}_B+\lambda\sqrt{\eta_1\eta_2(1-\eta_2)}(\ket{1}\bra{2}_B+\ket{2}\bra{1}_B)+
\left\{\eta_1-\lambda[\eta_1\eta_2+(1-\eta_1)(1-\eta_2)]\right\}\ket{2}\bra{2}_B,
\label{fmn}
\end{align}
and its largest eigenvalue, denoted by $\Lambda^{(110)}(\eta_1,\eta_2,\lambda)$, is 
\begin{align}
\Lambda^{(110)}(\eta_1,\eta_2,\lambda)=\frac{s(\eta_1,\eta_2,\lambda)+\sqrt{t(\eta_1,\eta_2,\lambda)}}{2}
\label{eq:Lambda}
\end{align}
with
\begin{align}
s(\eta_1,\eta_2,\lambda)&:=1-2(1-\eta_2)\lambda+
\eta_1(1+\lambda-2\eta_2\lambda)\in\mathbb{R},\\
t(\eta_1,\eta_2,\lambda)&:=
1+\eta_1\{\eta_1(1+\lambda-2\eta_2\lambda)^2-2
\left[1+\lambda-2\eta_2\lambda-2(1-\eta_2)\eta_2\lambda^2
\right]
\}>0.
\end{align}
As will be proven in Appendix~\ref{app:proLambda}, 
$\partial\Lambda^{(110)}(\eta_1,\eta_2,\lambda)/\partial\eta_1\ge0$ and 
$\partial\Lambda^{(110)}(\eta_1,\eta_2,\lambda)/\partial\eta_2\ge0$ hold for any $\lambda>0$ and any $\eta_1$ and $\eta_2$ 
of $0<\eta_1,\eta_2<1$, and hence $\Lambda^{(110)}(\eta_1,\eta_2,\lambda)$ is non-decreasing. 
Therefore, we have
\begin{align}
\Lambda^{(110)}(\eta_1,\eta_2,\lambda)\le \Lambda^{(110)}(\eta_1^U,\eta_2^U,\lambda).
\label{LAMUP}
\end{align}
Note that as defined in Eq.~(\ref{eq:symmetricrange}), $\eta_1^U$ and $\eta_2^U$ are the upper bounds on $\eta_1$ and $\eta_2$, respectively. \\
(ii) Second, we consider the case of  $\vec{a}=101$. $\hat{T}_{101}$ is written as
\begin{align}
\hat{T}_{101}=-\lambda(1-\eta_2)\ket{1}\bra{1}_B-\lambda\eta_2\ket{3}\bra{3}_B,
\label{fmn2}
\end{align}
which is negative for $\lambda>0$. 
\\
(iii) Finally, we consider the case of  $\vec{a}=011$. $\hat{T}_{011}$ is expressed as
\begin{align}
\hat{T}_{011}=\{1-\eta_1-\lambda[(1-\eta_1)(1-\eta_2)+\eta_1\eta_2]\}
\ket{2}\bra{2}_B+\lambda\sqrt{(1-\eta_1)(1-\eta_2)\eta_2}
(\ket{2}\bra{3}_B+\ket{3}\bra{2}_B)+(1-\lambda\eta_2)\ket{3}\bra{3}_B,
\end{align}
and its largest eigenvalue, denoted by $\Lambda^{(011)}(\eta_1,\eta_2,\lambda)$, 
and $\Lambda^{(110)}(\eta_1,\eta_2,\lambda)$ in Eq.~(\ref{eq:Lambda}) are related as
\begin{align}
\Lambda^{(011)}(\eta_1,\eta_2,\lambda)=\Lambda^{(110)}(1-\eta_1,1-\eta_2,\lambda).
\label{checkwant}
\end{align}
As a result of the discussion in (i)-(iii) above, Eq.~(\ref{kyoaya}) becomes
\begin{align}
&(\hat{U}_{AB}\hat{P}_1\hat{U}_{AB}^\dagger)(\hat{U}_{AB}\hat{e}_{\ph}(\eta_1)\hat{U}_{AB}^\dagger-\lambda
\hat{U}_{AB}\hat{e}_{\bit}(\eta_1,\eta_2)\hat{U}^\dagger_{AB})(\hat{U}_{AB}\hat{P}_1\hat{U}_{AB}^\dagger)\notag\\
\le&
\lambda|2\eta_2-1|+\hat{P}(\ket{110}_{\bm{A}})\otimes\hat{I}_B\Lambda^{(110)}(\eta_1,\eta_2,\lambda)
+\hat{P}(\ket{011}_{\bm{A}})\otimes\hat{I}_B
\Lambda^{(110)}(1-\eta_1,1-\eta_2,\lambda).
\label{eq:finalinter}
\end{align}
Under the assumption of the symmetric ranges for $\mathcal{R}_1$ and $\mathcal{R}_2$ 
with respect to 1/2 as described in Eq.~(\ref{eq:symmetricrange}), 
if $\eta_1\in \mathcal{R}_1$ and $\eta_2\in \mathcal{R}_1$, then $1-\eta_1\in \mathcal{R}_1$ and 
$1-\eta_2\in \mathcal{R}_1$ are satisfied. 
Therefore, from Eq.~(\ref{LAMUP}), both $\Lambda^{(110)}(\eta_1,\eta_2,\lambda)$ and 
$\Lambda^{(110)}(1-\eta_1,1-\eta_2,\lambda)$ in Eq.~(\ref{eq:finalinter}) 
are upper-bounded by $\Lambda^{(110)}(\eta_1^U,\eta_2^U,\lambda)$; this leads to 
\begin{align}
&(\hat{U}_{AB}\hat{P}_1\hat{U}_{AB}^\dagger)(\hat{U}_{AB}\hat{e}_{\ph}(\eta_1)\hat{U}_{AB}^\dagger-\lambda
\hat{U}_{AB}\hat{e}_{\bit}(\eta_1,\eta_2)\hat{U}^\dagger_{AB})(\hat{U}_{AB}\hat{P}_1\hat{U}_{AB}^\dagger)\notag\\
\le&
\lambda|2\eta_2-1|+[\hat{P}(\ket{110}_{\bm{A}})
+\hat{P}(\ket{011}_{\bm{A}})]\otimes\hat{I}_B\Lambda^{(110)}(\eta^U_1,\eta^U_2,\lambda).
\label{eq:finalinter2}
\end{align}
Below, we calculate $\lambda$ such that $\Lambda^{(110)}(\eta^U_1,\eta^U_2,\lambda)$ is equal to zero. 
From Eq.~(\ref{eq:Lambda}), the sufficient condition of $\Lambda^{(110)}(\eta^U_1,\eta^U_2,\lambda)=0$ is 
$s(\eta^U_1,\eta^U_2,\lambda)<0\wedge s(\eta^U_1,\eta^U_2,\lambda)^2=t(\eta^U_1,\eta^U_2,\lambda)$, and we have
\begin{align}
\begin{cases}
s(\eta^U_1,\eta^U_2,\lambda)<0\\
s(\eta^U_1,\eta^U_2,\lambda)^2=t(\eta^U_1,\eta^U_2,\lambda)
\end{cases}
\Longleftrightarrow
\begin{cases}
\lambda>\lambda^L:=(1+\eta_1^U)/\left[(1-\eta_1^U)(1-2\eta_2^U)+1\right]>0
\\
f(\lambda):=
(1-\eta_1^U)(1-\eta_2^U)^2\lambda^2-[1-(1-\eta_1^U)\eta_2^U]\lambda+\eta_1^U=0.
\end{cases}
\end{align}
The discriminant of the quadratic polynomial $f(\lambda)$ is non-negative for any $\eta_1^U$ and $\eta_2^U$ of 
$1/2\le\eta_1^U,\eta_2^U<1$, and the solutions of $f(\lambda)=0$ are given by
\begin{align}
\lambda=
\frac{1-(1-\eta_1^U)\eta_2^U\pm\sqrt{
[1-(1-\eta_1^U)\eta_2^U]^2-4\eta_1^U(1-\eta_1^U)(1-\eta_2^U)^2}}{2(1-\eta_1^U)(1-\eta_2^U)^2}
=:\lambda^{\pm}.
\end{align} 
It is straightforward to show $\lambda^+>\lambda^L$ for any $\eta_1^U$ and $\eta_2^U$, and hence 
$\Lambda^{(110)}(\eta^U_1,\eta^U_2,\lambda^+)=0$. 
By adopting this value $\lambda^+$ as the value of $\lambda$ in Eq.~(\ref{eq:finalinter2}), we finally obtain
\begin{align}
\hat{P}_1\hat{e}_{\ph}(\eta_1)\hat{P}_1
\le\lambda^+\left[|2\eta_2-1|+\hat{P}_1\hat{e}_{\bit}(\eta_1,\eta_2)\hat{P}_1\right]
\le\lambda^+\left[2\delta^{(\U{BS})}_2+\hat{P}_1\hat{e}_{\bit}(\eta_1,\eta_2)\hat{P}_1\right].
\end{align}
This ends the proof of Lemma~\ref{lemma1}. 

\section{Proof of Lemma~\ref{lemma2}}
\label{app:l2}
In this section, we prove Lemma~\ref{lemma2}. In Ref.~\cite{dps2019}, the statement in this lemma 
is proven only for the case of $\eta_1=\eta_2=0.5$. 
Our lemma generalizes it to hold for any $\eta_1$ and $\eta_2$ within the range of $0<\eta_1<1$ and $0<\eta_2<1$. 
We initiate the discussion by recognizing that we can conduct the two measurements 
$\hat{M}_1:=\{\hat{e}_\bit(\eta_1,\eta_2),\hat{I}_{\bm{A}B}-\hat{e}_\bit(\eta_1,\eta_2)\}$ and 
$\hat{M}_2:=\{\hat{P}_0+\hat{P}_2, \hat{P}_1+\hat{P}_3\}$ simultaneously 
because the outcomes of $\hat{M}_1$ and $\hat{M}_2$ are obtained by measuring different systems. 
For instance, when a detection event occurs at TS1, the outcomes of $\hat{M}_1$ and $\hat{M}_2$ are obtained 
by measuring systems $A_1$ and $A_2$, respectively. 
The simultaneous measurability of $\hat{M}_1$ and $\hat{M}_2$ 
is equivalent to $[\hat{e}_\bit(\eta_1,\eta_2),\hat{P}_0+\hat{P}_2]=0$, and 
using this commutation relation along with the fact that $\hat{P}_i$ is a projector, we have
\begin{align}
\hat{e}_{\bit}(\eta_1,\eta_2)=(\hat{P}_1+\hat{P}_3)\hat{e}_{\bit}(\eta_1,\eta_2)(\hat{P}_1+\hat{P}_3)+
(\hat{P}_0+\hat{P}_2)\hat{e}_{\bit}(\eta_1,\eta_2)(\hat{P}_0+\hat{P}_2).
\label{eq:intern}
\end{align}
By employing $(\hat{P}_0+\hat{P}_2)\hat{e}_{\bit}(\eta_1,\eta_2)(\hat{P}_0+\hat{P}_2)\ge0$ and 
$\hat{P}_3\hat{e}_{\bit}(\eta_1,\eta_2)\hat{P}_3\ge0$, Eq.~(\ref{eq:intern}) leads to
\begin{align}
\hat{P}_1\hat{e}_{\bit}(\eta_1,\eta_2)\hat{P}_1\le\hat{e}_{\bit}(\eta_1,\eta_2)-
(\hat{P}_1\hat{e}_{\bit}(\eta_1,\eta_2)\hat{P}_3+\hat{P}_3\hat{e}_{\bit}(\eta_1,\eta_2)\hat{P}_1).
\label{eq:ind2}
\end{align}
Since $|\tr\hat{O}|=|\tr\hat{O}^\dagger|$ is satisfied for any square matrix $\hat{O}$, we have from Eq.~(\ref{eq:ind2}) that 
\begin{align}
\tr[\hat{P}_1\hat{e}_{\bit}(\eta_1,\eta_2)\hat{P}_1\hat{\sigma}]
&\le\tr[\hat{e}_{\bit}(\eta_1,\eta_2)\hat{\sigma}]-\tr[(\hat{P}_1\hat{e}_{\bit}(\eta_1,\eta_2)\hat{P}_3+
\hat{P}_3\hat{e}_{\bit}(\eta_1,\eta_2)\hat{P}_1)\hat{\sigma}]\\
&\le\tr[\hat{e}_{\bit}(\eta_1,\eta_2)\hat{\sigma}]+2|\tr[\hat{P}_1\hat{e}_{\bit}(\eta_1,\eta_2)\hat{P}_3\hat{\sigma}]|
\label{reghjajyfth}
\end{align}
holds for any state $\hat{\sigma}$ of systems $\bm{A}B$. 
By defining $\hat{G}:=\hat{P}_3\hat{\sigma}\hat{P}_1$ and $\hat{T}:=2\hat{P}_1\hat{e}_{\bit}(\eta_1,\eta_2)\hat{P}_3$, 
the second term of the right-hand side of Eq.~(\ref{reghjajyfth}) is equal to $|\tr(\hat{T}\hat{G})|$, and H\"{o}lder's inequality 
leads to
\begin{align}
|\tr(\hat{T}\hat{G})|\le||\hat{T}||_\infty||\hat{G}||_1.
\label{eq:Gfinish}
\end{align}
Here, $||\hat{O}||_1$ denotes the Schatten-1 norm given by 
$||\hat{O}||_1=\tr |\hat{O}|=\tr\sqrt{\hat{O}^\dagger \hat{O}}$, and $||\hat{O}||_{\infty}$ represents the operator norm 
defined as the minimum value of $c\ge0$ satisfying $||\hat{O}\ket{v}||/||\ket{v}||\le c$ for any vector $\ket{v}$. 

In the following, we derive respective upper bounds on $||\hat{T}||_\infty$ and $||\hat{G}||_1$. 
As for $||\hat{G}||_1$, it can be rewritten with a unitary operator $\hat{W}$ as $||\hat{G}||_1=|\tr(\hat{G}\hat{W})|$, and 
using the Cauchy-Schwarz inequality gives its upper bound as
\begin{align}
|\tr(\hat{G}\hat{W})|=|(\sqrt{\hat{\sigma}}\hat{P}_3,\sqrt{\hat{\sigma}}\hat{P}_1\hat{W})|
\le\sqrt{
(\sqrt{\hat{\sigma}}\hat{P}_3,\sqrt{\hat{\sigma}}\hat{P}_3)(\sqrt{\hat{\sigma}}\hat{P}_1\hat{W},\sqrt{\hat{\sigma}}\hat{P}_1\hat{W})
}
=\sqrt{\tr(\hat{P}_3\hat{\sigma})}\sqrt{\tr(\hat{P}_1\hat{\sigma})}.
\label{eq:g1upper}
\end{align}
Our remaining task to complete the proof of Lemma~\ref{lemma2} is then to prove $||\hat{T}||_\infty\le1$. 
By substituting the definitions in Eqs.~(\ref{koret1}), (\ref{eq:defP1}) and (\ref{eq:exp3}) to $\hat{T}=2\hat{P}_1\hat{e}_{\bit}(\eta_1,\eta_2)\hat{P}_3$, we have
\begin{align}
\hat{T}=\ket{001}\bra{111}_{\bm{A}}\otimes\underbrace{(\hat{\Pi}_{1,D_1}-\hat{\Pi}_{1,D_0})}_{=:\hat{X}}
+\ket{100}\bra{111}_{\bm{A}}\otimes\underbrace{(\hat{\Pi}_{2,D_1}-\hat{\Pi}_{2,D_0})}_{=:\hat{Y}}, 
\end{align}
and its Gram matrix is given by 
\begin{align}
\hat{T}^\dagger \hat{T}=\ket{111}\bra{111}_{\bm{A}}\otimes(\hat{X}^2+\hat{Y}^2).
\label{gekhjkhj}
\end{align}
By using the expressions in Eqs.~(\ref{ertjklghertjgyh})-(\ref{bobpovm4}), we obtain
\begin{align}
&\hat{X}=\underbrace{(2\eta_2-1)}_{=:p}\ket{1}\bra{1}_B+
\underbrace{\eta_1(1-2\eta_2)}_{=:q}\ket{2}\bra{2}_B-\underbrace{2\sqrt{\eta_1\eta_2(1-\eta_2)}}_{=:r}(\ket{1}\bra{2}_B+\ket{2}\bra{1}_B),\\
&\hat{X}^2=(p^2+r^2)\ket{1}\bra{1}_B+(q^2+r^2)\ket{2}\bra{2}_B-r(p+q)(\ket{1}\bra{2}_B+\ket{2}\bra{1}_B),\\
&\hat{Y}=\underbrace{(2\eta_2-2\eta_1\eta_2-1+\eta_1)}_{=:a}\ket{2}\bra{2}_B+
\underbrace{(1-2\eta_2)}_{=:b}\ket{3}\bra{3}_B-\underbrace{2\sqrt{\eta_2(1-\eta_1)(1-\eta_2)}}_{=:c}(\ket{2}\bra{3}_B+\ket{3}\bra{2}_B),\\
&\hat{Y}^2=(a^2+c^2)\ket{2}\bra{2}_B+(b^2+c^2)\ket{3}\bra{3}_B-c(a+b)(\ket{2}\bra{3}_B+\ket{3}\bra{2}_B).
\end{align}
Substituting $\hat{X}^2$ and $\hat{Y}^2$ to Eq.~(\ref{gekhjkhj}) yields
\begin{align}
\hat{T}^\dagger\hat{T}=\ket{111}\bra{111}_{\bm{A}}\otimes\hat{M}_B,
\label{eq:matrixMB}
\end{align}
where $\hat{M}_B$ is represented in the basis $\mathcal{B}=\{\ket{1}_B,\ket{2}_B,\ket{3}_B\}$ as
\begin{align}
\hat{M}_B=
\begin{pmatrix}
1-4\eta_2(1-\eta_1)(1-\eta_2)& -2(1-\eta_1)(2\eta_2-1)\sqrt{\eta_1\eta_2(1-\eta_2)} &0\\
-2(1-\eta_1)(2\eta_2-1)\sqrt{\eta_1\eta_2(1-\eta_2)} &1-2(1-\eta_1)\eta_1(1-2\eta_2)^2 
&-2\eta_1(1-2\eta_2)\sqrt{(1-\eta_1)(1-\eta_2)\eta_2} \\
0&-2\eta_1(1-2\eta_2)\sqrt{(1-\eta_1)(1-\eta_2)\eta_2}& 1-4(1-\eta_2)\eta_1\eta_2
\end{pmatrix}.
\end{align}
The eigenvalues of $\hat{M}_B$ 
are 1 and $\gamma_{\pm}(\eta_1,\eta_2)=1-a(\eta_1,\eta_2)\pm\sqrt{b(\eta_1,\eta_2)}$ 
with
\begin{align}
a(\eta_1,\eta_2):=&(1-2\eta_2)^2\eta_1(1-\eta_1)+2\eta_2(1-\eta_2)>0,\\
b(\eta_1,\eta_2):=&-2\eta_1^3(1-2\eta_2)^4+\eta_1^4(1-2\eta_2)^4+4(1-\eta_2)^2\eta_2^2-16\eta_1(1-\eta_2)^2\eta_2^2+\eta_1^2\left[1-8(1-2\eta_2)^2(1-\eta_2)\eta_2\right]\ge0.
\label{kkefe}
\end{align}
It is straightforward to see that $a(\eta_1,\eta_2)^2-b(\eta_1,\eta_2)=4\eta_1\eta_2(1-\eta_1)(1-\eta_2)\ge0$ holds, and hence $\gamma_+(\eta_1,\eta_2)\le1$. 
Substituting $\hat{M}_B\le \hat{I}_B$ to Eq.~(\ref{eq:matrixMB}) gives 
$\hat{T}^\dagger \hat{T}\le \hat{I}_{\bm{A}B}$, which results in $||\hat{T}||_\infty\le1$. 
This ends the proof of Lemma~\ref{lemma2}.

\section{Proof of $\frac{\partial\Lambda^{(110)}(\eta_1,\eta_2,\lambda)}{\partial\eta_1}\ge0$ 
and $\frac{\partial\Lambda^{(110)}(\eta_1,\eta_2,\lambda)}{\partial\eta_2}\ge0$
}\label{app:proLambda}
In this section, we prove that 
\begin{align}
\Lambda^{(110)}(\eta_1,\eta_2,\lambda)=\frac{s(\eta_1,\eta_2,\lambda)+\sqrt{t(\eta_1,\eta_2,\lambda)}}{2}
\end{align}
with
\begin{align}
s(\eta_1,\eta_2,\lambda)&:=1-2(1-\eta_2)\lambda+
\eta_1(1+\lambda-2\eta_2\lambda)\in\mathbb{R},\\
t(\eta_1,\eta_2,\lambda)&:=
1+\eta_1\{\eta_1(1+\lambda-2\eta_2\lambda)^2-2
\left[1+\lambda-2\eta_2\lambda-2(1-\eta_2)\eta_2\lambda^2
\right]
\}>0
\end{align}
is non-decreasing with respect to $\eta_1$ and $\eta_2$. 

\subsection{
Proof of $\frac{\partial\Lambda^{(110)}(\eta_1,\eta_2,\lambda)}{\partial\eta_1}\ge0$
}
In the function $\Lambda^{(110)}(\eta_1,\eta_2,\lambda)$, we consider $\eta_1$ as a variable and $\eta_2$ 
as a constant value. The partial derivative with respect to $\eta_1$ is calculated as follows:
\begin{align}
\frac{\partial\Lambda^{(110)}(\eta_1,\eta_2,\lambda)}{\partial \eta_1}
=\frac{A\sqrt{t(\eta_1,\eta_2,\lambda)}+B}
{2\sqrt{t(\eta_1,\eta_2,\lambda)}}
\label{eq:targeteta1}
\end{align}
with
\begin{align}
A:=1+\lambda-2\lambda \eta_2,~
B:=\eta_1(1+\lambda-2\eta_2\lambda)^2+\lambda\left[2\eta_2(1+\lambda-\lambda\eta_2)-1\right]-1.
\label{eqST}
\end{align}
The target inequality that we aim to prove is
\begin{align}
\frac{\partial \Lambda^{(110)}(\eta_1,\eta_2,\lambda)}{\partial \eta_1}\ge0,
\label{eq:targetLeta1}
\end{align}
and we show this by considering the four cases according to the signs of $A$ and $B$.

\subsubsection{Case of $A\le0$ and $B\le0$}
First, we demonstrate that $A\le0$ and $B\le0$ do not hold simultaneously. 
$A\le0$ is equivalent to 
\begin{align}
1<\lambda~\U{and}~\frac{\lambda+1}{2\lambda}\le \eta_2<1,
\label{eq:con}
\end{align}
and under this condition we show $B>0$. 
By removing the first term of $B$ in Eq.~(\ref{eqST}), we have 
$B\ge \lambda\left[2\eta_2(1+\lambda-\lambda\eta_2)-1\right]-1$. 
Then, it suffices to show that this lower bound is positive, which is equivalent to
\begin{align}
g(\eta_2):=\left(\eta_2-\frac{\lambda+1}{2\lambda}\right)^2+\frac{\lambda+1}{2\lambda^2}-\frac{(\lambda+1)^2}{4\lambda^2}
<0.
\label{eq:Fgap}
\end{align}
Under Eq.~(\ref{eq:con}), $g(\eta_2)$ is upper bounded by $g(1)$, and we have 
$g(\eta_2)<g(1)=(1-\lambda)/2\lambda^2<0$ for $1<\lambda$.

\subsubsection{Case of $A\ge0$ and $B\ge0$}
If $A\ge0$ and $B\ge0$, Eq.~(\ref{eq:targetLeta1}) trivially holds. 

\subsubsection{Case of $A\ge0$ and $B\le0$}
If $A\ge0$ and $B\le0$, it suffices to show $(A\sqrt{t(\eta_1,\eta_2,\lambda)})^2\ge B^2$, which is equivalent to
\begin{align}
\label{defFF}
f(\eta_2):=
\left[
\eta_2-\frac{\lambda+2}{2\lambda}
\right]^2+\frac{\lambda+1}{\lambda^2}-\frac{(\lambda+2)^2}{4\lambda^2}
\ge0.
\end{align}
Direct calculation reveals that
\begin{align}
A\ge0~\U{and}~B\le0\Longleftrightarrow&\U{(I)~or~(II)}
\end{align}
with
\begin{align}
(\U{I}): 
0<\lambda\le1~\wedge~\left[\left(0<\eta_1\le\frac{1}{1+\lambda}\right)~\lor~
\left(\frac{1}{1+\lambda}<\eta_1<1~\wedge~
\frac{1+\lambda}{2\lambda}-\frac{1}{2}\sqrt{\frac{1-\lambda^2}{(2\eta_1-1)\lambda^2}}\le \eta_2<1\right)
\right]
\end{align}
\begin{align}
(\U{II}): 
1<\lambda~\wedge~0<\eta_1\le\frac{1}{1+\lambda}~\wedge~0<\eta_2\le
\frac{1+\lambda}{2\lambda}-\frac{1}{2}\sqrt{\frac{\lambda^2-1}{(1-2\eta_1)\lambda^2}}=:\eta_2^{\max},
\end{align}
and
we prove $f(\eta_2)\ge0$ for both cases of (I) and (II). 

In case (I), the axis of symmetry, $(\lambda+2)/2\lambda$, of the quadratic function $f(\eta_2)$ is larger than 1 due to $0<\lambda\le1$. Therefore, we obtain $f(\eta_2)\ge f(1)=(1-\lambda)/\lambda^2\ge0$ for $0<\lambda\le1$. 

In case (II), as $\eta_2^{\max}$ is smaller than the axis of symmetry, $(\lambda+2)/2\lambda$, of the quadratic function $f(\eta_2)$, 
we have $f(\eta_2)\ge f(\eta_2^{\max})$. Since
\begin{align}
\eta_2^{\max}\le E:=\frac{1+\lambda}{2\lambda}-\frac{1}{2}\sqrt{\frac{\lambda^2-1}{\lambda^2}}<\frac{\lambda+2}{2\lambda},
\end{align}
we obtain $f(\eta_2^{\max})\ge f(E)=\frac{\sqrt{\lambda^2-1}}{2\lambda^2}\ge0$ for $\lambda>1$. 
Therefore, we conclude $f(\eta_2)\ge0$ for both cases of (I) and (II). 

\subsubsection{Case of $A\le0$ and $B\ge0$}
If $A\le0$ and $B\ge0$, it suffices to show $B^2\ge(A\sqrt{t(\eta_1,\eta_2,\lambda)})^2$, which is equivalent to 
$f(\eta_2)\le0$. Note that $f(\eta_2)$ is defined in Eq.~(\ref{defFF}). 
We need to show the maximum of $f(\eta_2)$ is upper bounded by zero under the following condition:
\begin{align}
A\le0~\U{and}~B\ge0\Longleftrightarrow
1<\lambda~\wedge~\eta_2^{\min}:=\frac{1+\lambda}{2\lambda}\le \eta_2<1.
\end{align}
The maximum of $f(\eta_2)$ is either $f(\eta_2^{\min})=(1-\lambda^2)/4\lambda^2$ or 
$f(1)=(1-\lambda)/\lambda^2$, and both values are negative for $1<\lambda$. 

Combining the results of the above four cases results in Eq.~(\ref{eq:targetLeta1}). 

\subsection{Proof of $\frac{\partial\Lambda^{(110)}(\eta_1,\eta_2,\lambda)}{\partial\eta_2}\ge0$}
In the function $\Lambda^{(110)}(\eta_1,\eta_2,\lambda)$, we consider $\eta_2$ as a variable and $\eta_1$ as a 
constant value. The partial derivative with respect to $\eta_2$ is calculated as follows:
\begin{align}
\frac{\partial \Lambda^{(110)}(\eta_1,\eta_2,\lambda)}
{\partial \eta_2}=\frac{C\sqrt{t(\eta_1,\eta_2,\lambda)}+D}{\sqrt{t(\eta_1,\eta_2,\lambda)}}
\end{align}
with
\begin{align}
C:=\lambda(1-\eta_1)>0,~
D:=(1-\eta_1)\eta_1\lambda[1+(1-2\eta_2)\lambda].
\end{align}
If $D\ge0$, it is trivial that the target inequality
\begin{align}
\frac{\partial \Lambda^{(110)}(\eta_1,\eta_2,\lambda)}{\partial \eta_2}\ge0
\label{eq:targetLeta2}
\end{align}
holds, and hence in the following we consider the case of $D\le0$, which is equivalent to
\begin{align}
D\ge0\Longleftrightarrow
\lambda>1~\wedge~\frac{\lambda+1}{2\lambda}\le \eta_2<1.
\label{eq6}
\end{align}
Then, for deriving Eq.~(\ref{eq:targetLeta2}), it suffices to show 
\begin{align}
(C\sqrt{t(\eta_1,\eta_2,\lambda)})^2\ge D^2
\Longleftrightarrow
g(\eta_2)\le\frac{1}{4\lambda^2\eta_1}
\label{eq:Leta2in}
\end{align}
under Eq.~(\ref{eq6}). Here, $g(\eta_2)$ is defined in Eq.~(\ref{eq:Fgap}). 
From the second condition in Eq.~(\ref{eq6}), we see that the maximum of $g(\eta_2)$ is upper bounded by $g(1)$, that is, 
$g(\eta_2)\le g(1)=(1-\lambda)/2\lambda^2$. 
From the first condition in Eq.~(\ref{eq6}), this upper bound is negative, and therefore Eq.~(\ref{eq:Leta2in}) holds.

\section{Security of DPS protocol with threshold detectors}
\label{app:thde}
\begin{figure*}[t]
\includegraphics[width=11cm]{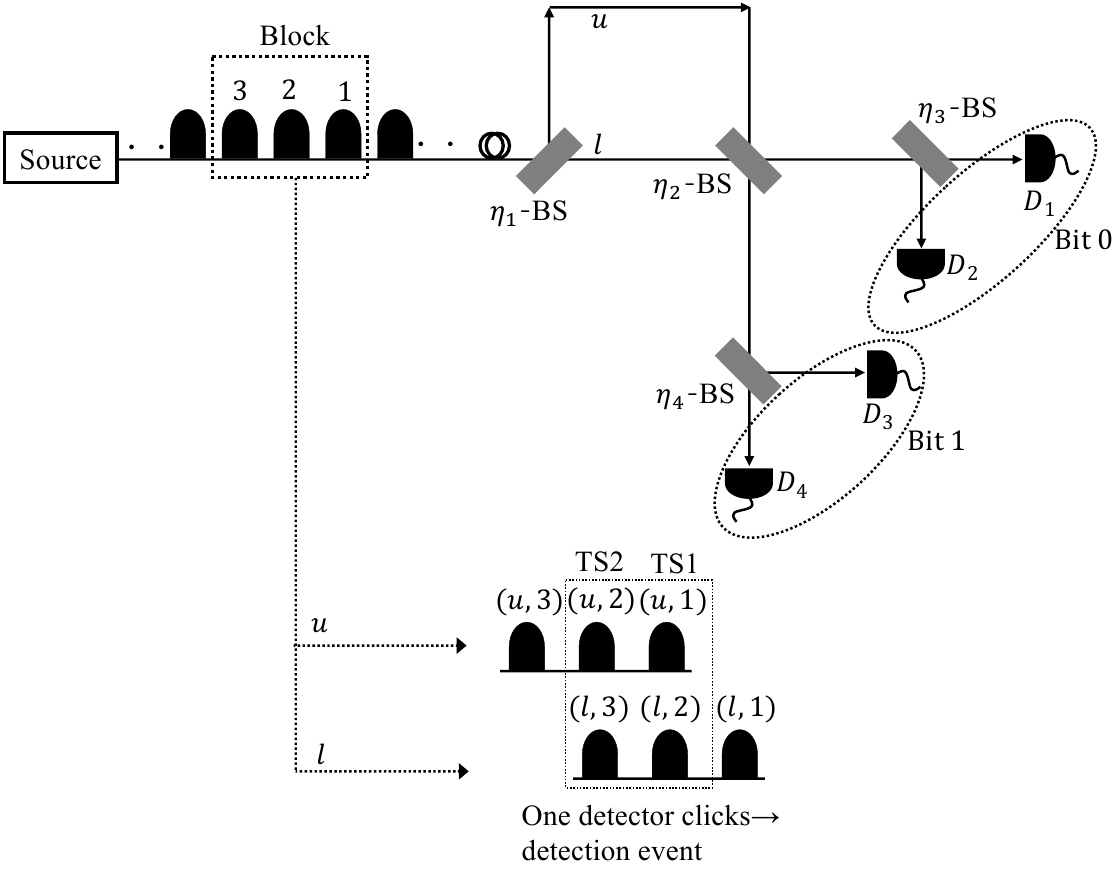}
\caption{
Experimental setup for our DPS protocol with threshold detectors. Alice sends blocks composed 
of three pulses to Bob, and he receives them with the one-bit
delay Mach-Zehnder interferometer and four threshold detectors $D_1, D_2, D_3$ and $D_4$. 
Here, $D_1$ and $D_2$ are detectors for reporting bit 0, while $D_3$ and $D_4$ detectors for reporting bit 1. 
The transmittance $\eta_1, \eta_2, \eta_3$ and $\eta_4$ of Bob's beam splitters 
can take any value within the ranges as stated in Eq.~(\ref{eq:symmetricrange}) and (\ref{assueta34}). 
$u$ and $l$ respectively represent the upper and lower arms of the Mach-Zehnder interferometer, 
and the pulse pairs $(u,1)$ and $(l,2)$, and $(u,2)$ and $(l,3)$ interfere at the BS2, and TS1 (TS2) 
is the time slot of detection of the first (second) pulse pair. 
A detection event occurs when only one detector clicks among the time slots TS1 and TS2. 
}
 \label{fig:actual2}
\end{figure*}
In this appendix, we present the security proof of our DPS protocol using threshold detectors at Bob's measurement unit. 
We begin by modifying Bob's device assumption~\ref{ass:b2} and our DPS protocol. 
Subsequently, we provide the security proof for the modified DPS protocol. 
The crux of our modified protocol is to use four threshold detectors instead of two 
and monitor the occurrences of multiple-click events, whose idea 
is presented in the security proof of the round-robin DPS protocol~\cite{rrdpssa}.
\\
\subsection{Modified Bob's assumption and protocol description}
\label{app:mbs}
The assumption~\ref{ass:b2} on Bob's measurement unit and the procedures of our DPS protocol 
are modified as follows (modifications are shown in bold font). 
\begin{enumerate}[label=(B\arabic*)]
\item
Bob employs {\bf four threshold detectors \bm{$D_1, D_2, D_3$} and \bm{$D_4$} 
that discriminate between the vacuum and a single or more photons} of a specific single optical mode. 
The detection inefficiency is modeled as a beam splitter (BS) followed by an ideal detector with a unit quantum efficiency. 
The quantum efficiencies are identical for {\bf four threshold detectors} and are denoted by $\eta_{\rm det}$. 
Moreover, we assume that the dark counting of the detector is simulated by a stray photon source positioned 
in front of Bob's measurement unit. 
\end{enumerate}
In front of $D_1$ and $D_2$ ($D_3$ and $D_4$), there is a beam splitter with transmittance $\eta_3$ ($\eta_4$)
to split the incoming light. 
As with assumption~\ref{ass:b1}, we assume that Alice and Bob do not know the exact transmittance but its ranges:
\begin{align}
\eta_3\in [1/2-\delta^{(\U{BS})}_3,1/2+\delta^{(\U{BS})}_3]~\U{and}~
\eta_4\in [1/2-\delta^{(\U{BS})}_4,1/2+\delta^{(\U{BS})}_4]
\label{assueta34}
\end{align}
with $0\le\delta^{(\U{BS})}_3,\delta^{(\U{BS})}_4<1/2$. For later convenience, we define 
$\delta^{(\rm BS)}:=\max\{\delta^{(\rm BS)}_3,\delta^{(\rm BS)}_4\}$. 
\\

Next, our DPS protocol with threshold detectors runs as follows (see Fig.~\ref{fig:actual2}). 
\begin{enumerate}
\item
Alice and Bob respectively execute the following steps (a) and (b) $N_{\em}$ times. 
\begin{enumerate}
\item
Alice uniformly and randomly chooses three bits $b_1b_2b_3\in\{0,1\}^3$, and according to the chosen bits, 
she sends state 
$\hat{\rho}_{S_1S_2S_3}^{b_1b_2b_3}$ of a single block to Bob via a quantum channel. 
\item
Bob splits the incoming three pulses into two pulse trains using the first BS (BS1). 
The $i$th pulse with $i\in\{1,2,3\}$ passing through the lower and upper arms of the Mach-Zehnder interferometer 
are labeled by $(l,i)$ and $(u,i)$, respectively. 
The pulse pairs $(u,1)$ and $(l,2)$, and $(u,2)$ and $(l,3)$ interfere at the second BS (BS2). 
We define the time slots of detection of the first and second pulse pairs as TS1 and TS2, respectively. 
We define a ``detection event" as the one in which {\bf only one detector clicks} in total in TS1 and TS2. 
The detection 
event at TS$j$ (with $j\in\{1,2\}$) determines the raw key bit $d$ depending on which of the {\bf four} detectors clicks. 
{\bf We also define the multi-click event as the one where detectors click two or more times in TS1 and TS2.}
\end{enumerate}
\item
{\bf Bob takes notes of the number of multi-click events \bm{$N_{\rm multi}$}.} 
Bob defines the set of detection events $\mathcal{S}\subset \{1,...,N_{\em}\}$ with length 
$|\mathcal{S}|:=N_{\det}$, the set of time slots at which Bob obtained the detection event, 
 i.e., $\{\U{TS}j_i\}_{i\in\mathcal{S}}$, and the raw key bits $(d_i)_{i\in\mathcal{S}}$. 
Here, $j_i$ and $d_i$ ($i\in\mathcal{S}$) respectively denote the values of $j$ and $d$ of the $i$th detection event. 
Within the detection events, 
Bob randomly assigns each detection event to a code event with probability $t$ or a sample event with probability 
$1-t$ (where $0<t<1$). 
Then, he obtains the code set $\mathcal{S}_{\code}$ with length $|\mathcal{S}_\code|:=N_{\code}$, 
the sample set $\mathcal{S}_{\sample}$ with length $|\mathcal{S}_\code|:=N_{\code}$, 
his sifted key $\kappa_B:=(d_i)_{i\in\mathcal{S}_{\code}}$, and the sample bit sequence 
$\kappa^{\sample}_B:=(d_i)_{i\in\mathcal{S}_{\sample}}$. 
\item
Bob announces $\mathcal{S}_{\code}, \mathcal{S}_{\sample}$, $\{\U{TS}j_i\}_{i\in\mathcal{S}}$ and 
$\kappa_B^\sample$ via an authenticated public channel.
\item
Alice obtains her sifted key $\kappa_A:=(b_{j_i}\oplus b_{j_i+1})_{i\in\mathcal{S}_\code}$ 
and the sample bit sequence $\kappa^{\sample}_A:=(b_{j_i}\oplus b_{j_i+1})_{i\in\mathcal{S}_\sample}$. 
\item
Alice estimates the bit error rate in the code events from the bit error rate in the sample events, selects a bit error correction code, and sends the syndrome information of her sifted key $\kappa_A$ to Bob by consuming pre-shared secret key 
of length $N_{\U{EC}}$. 
Using the syndrome information, Bob corrects bit errors in his sifted key and obtains the reconciled key 
$\kappa^{\rm rec}_B$. 
\item
Alice and Bob execute privacy amplification to respectively shorten $\kappa_A$ and $\kappa_B^{\U{rec}}$ by 
$N_\U{PA}$ to obtain their final keys of length $N_{\code}-N_\U{PA}$. 
\end{enumerate}
After the execution of the protocol, the net length of the increased secret key is given by
\begin{align}
\ell=N_{\code}-N_{\U{PA}}-N_{\U{EC}}.
\label{eq:keyrateTh}
\end{align}
For later use, we define the following parameter
\begin{align}
N_{\bit}:=\wt(\kappa_A^{\sample}\oplus\kappa_B^{\sample}).
\end{align}
\subsection{Security proof of DPS protocol with threshold detectors}
For the DPS protocol described in Sec.~\ref{app:mbs}, the result of its security proof is stated as follows. 
\begin{theorem}
\label{th:keyrateTh}
If Alice and Bob shorten their reconciled keys of length $N_{\U{code}}$ by
\begin{align}
&N_{\U{PA}}=\frac{N_{\rm multi}}{(1+2\delta^{(\rm BS)})(1/2-\delta^{(\rm BS)})}+
\left(N_{\code}-\frac{N_{\rm multi}}{(1+2\delta^{(\rm BS)})(1/2-\delta^{(\rm BS)})}\right)\times\notag\\
&h\left(\frac{t\lambda(\eta_1^U,\eta_2^U)\left(\frac{N_{\bit}}{1-t}+
N_{\em}\sqrt{q_1q_3}+2\delta^{(\U{BS})}_2N_{\rm det}\right)+tN_{\em}q_2}
{
N_{\rm code}-N_{\rm multi}/[(1+2\delta^{(\rm BS)})(1/2-\delta^{(\rm BS)})]
}
\right)
\label{eq:moPA}
\end{align}
in the privacy amplification step, they share a secret key of length
\begin{align}
\ell=N_{\code}-N_{\U{PA}}-N_{\U{EC}}.
\end{align}
Note that all the parameters appearing on the right-hand side of Eq.~(\ref{eq:moPA}) can be obtained in the actual 
experiment
\footnote{
Here, $N_{\rm multi}, \delta^{(\rm BS)}, N_{\code}, N_{\rm det}, N_{\rm em}, t$ and $N_{\rm bit}$ are introduced in Sec.~\ref{app:mbs}, and $\delta^{(\rm BS)}_2, \eta_1^U, \eta_2^U$ and $q_n$ are defined in Sec.~\ref{sec:deviceass}.
}. Also, $h(x)$ is the binary entropy function, and $\lambda(x,y)$ is defined in Eq.~(\ref{def:lambdade}).
\end{theorem}

\begin{figure*}[t]
\includegraphics[width=9cm]{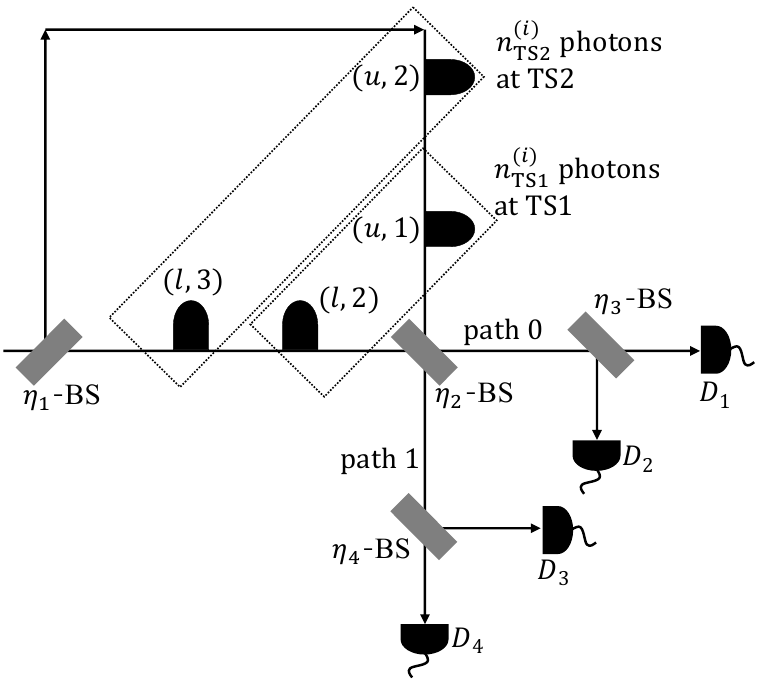}
\caption{
Bob's virtual measurement to learn the number of photons $n^{(i)}_{\rm TS1}$ and $n^{(i)}_{\rm TS2}$ 
contained in the pulses at TS1 and TS2 for the $i$th incoming block, respectively. 
If the total number of photons at TS1 and TS2 is two or more, namely, 
$n^{(i)}_{\rm TS1,2}:=n^{(i)}_{\rm TS1}+n^{(i)}_{\rm TS2}\ge2$, 
the multi-click event always occurs except when all the photons are present at TS1 or TS2 and path~0 or path~1. 
Here, ``path 0" (``path 1") is the output path of the Mach-Zehnder interferometer outputting bit 0 (1). 
}
 \label{fig:app}
\end{figure*}
{\it Proof of Theorem~\ref{th:keyrateTh}.} 
For deriving the amount of privacy amplification $N_{\U{PA}}$, it is convenient to introduce a virtual Bob's measurement 
to learn the total number of photons
\begin{align}
n^{(i)}_{\rm TS1,2}:=n^{(i)}_{\rm TS1}+n^{(i)}_{\rm TS2}
\label{nts12}
\end{align}
contained in the pulses at TS1 and TS2 
for each $i$th incoming block with $1\le i\le N_{\rm em}$. Here, 
$n^{(i)}_{\rm TS1}$ and $n^{(i)}_{\rm TS2}$ denote the number of photons contained in the pulses 
at TS1 and TS2, respectively (see Fig.~\ref{fig:app}). 
We can introduce this virtual measurement as it never changes the statistics of actual Bob's measurement. 
By this measurement, we can classify each incoming block into untagged events with $n^{(i)}_{\rm TS1,2}=1$ 
and tagged events with $n^{(i)}_{\rm TS1,2}\ge2$, and 
privacy amplification can be executed separately for these events. 
In the following discussions, we define the amount of privacy amplification for the tagged and untagged events as 
$N^{\rm tag}_{\rm PA}$ and $N^{\rm untag}_{\rm PA}$, respectively. Also, we define the number of untagged code events 
as $N_{\rm code}^{\rm untag}$.

Considering the worst-case scenario where the information of the sifted key generated from the tagged events is completely leaked to Eve, we obtain
\begin{align}
N^{\rm tag}_{\rm PA}+N^{\rm untag}_{\rm PA}\le (N_{\rm code}-N_{\rm code}^{\rm untag})+N_{\rm code}^{\rm untag}
h\left(\frac{N_{\rm ph}^{\rm code,untag}}{N_{\rm code}^{\rm untag}}\right),
\label{nthl}
\end{align}
where $N_{\rm ph}^{\rm code,untag}$ denotes the number of phase errors associated with the untagged code events. 
The untagged code events correspond to the code events of the DPS protocol with 
photon-number-resolving detectors (see Sec.~\ref{sec:proto}), and 
the number of phase errors for these events was already derived in Eq.~(\ref{eq:derapT}). 
Therefore, directly borrowing from Eq.~(\ref{eq:derapT}), we have
\begin{align}
N_{\rm ph}^{\rm code,untag}\le
t\lambda(\eta_1^U,\eta_2^U)\left(\frac{N_{\bit}}{1-t}+
N_{\em}\sqrt{q_1q_3}+2\delta^{(\U{BS})}_2N_{\rm det}\right)+tN_{\em}q_2.
\end{align}
Since the right-hand side of Eq.~(\ref{nthl}) is monotonically decreasing with respect to $N_{\rm code}^{\rm untag}$, 
the remaining task to complete the proof of Theorem~\ref{th:keyrateTh} is to derive the lower bound on 
$N_{\rm code}^{\rm untag}$. 
For this, we observe that
\begin{align}
N_{\rm code}^{\rm untag}\ge N_{\rm code}-N_{\rm tag}
\label{eq:g9}
\end{align}
holds, where $N_{\rm tag}:=|\{1\le i\le N_{\rm em}|n^{(i)}_{\rm TS1,2}\ge2\}|$ denotes the number of tagged events in which 
multiple photons are contained in the pulses at TS1 and TS2. 
Although $N_{\rm tag}$ cannot be directly observed in the actual experiment, we can derive its upper bound using the number of multi-click events $N_{\rm multi}$ as 
\begin{align}
N_{\rm tag}\le
\frac{1}{(1+2\max\{\delta^{(\rm BS)}_3,\delta^{(\rm BS)}_4\})(1/2-\max\{\delta^{(\rm BS)}_3,\delta^{(\rm BS)}_4\})}N_{\rm multi}.
\label{l:keyrateTh}
\end{align}
By substituting this upper bound into Eq.~(\ref{eq:g9}), substituting the resulting lower bound on $N_{\rm code}^{\rm untag}$ 
into Eq.~(\ref{nthl}), and regarding the resulting upper bound on Eq.~(\ref{nthl}) as $N_{\rm PA}$, we finally obtain 
Theorem~\ref{th:keyrateTh}.
For completeness of this paper, we prove Eq.~(\ref{l:keyrateTh}) below. 
\\\\
{\it Proof of Eq.~(\ref{l:keyrateTh}).} 
To derive the upper bound on the number $N_{\rm tag}$ of tagged events, 
we consider probabilistic trials of measuring the total number of photons $n^{(i)}_{\rm TS1,2}$ at TS1 and TS2 and 
determining whether the multi-click event occurs or not. 
Let $y_i$ denote the result of which of the four detectors clicks for the $i$th incoming block, and with this definition, 
the probability of observing the tagged and multi-click event for the $i$th incoming block 
conditioned on the previous outcomes $\bm{y}_{i-1}:=y_1...y_{i-1}$ and 
$\bm{n^{(i-1)}_{\rm TS1,2}}:=n^{(1)}_{\rm TS1,2}...n^{(i-1)}_{\rm TS1,2}$
is written as
\begin{align}
&\U{Pr}[y_i={\rm multi~click}, n^{(i)}_{\rm TS1,2}\ge2|\bm{y}_{i-1},\bm{n^{(i-1)}_{\rm TS1,2}}]=
\U{Pr}[y_i={\rm multi~click}|n^{(i)}_{\rm TS1,2}\ge2,\bm{y}_{i-1},\bm{n^{(i-1)}_{\rm TS1,2}}]
\U{Pr}[n^{(i)}_{\rm TS1,2}\ge2,\bm{y}_{i-1},\bm{n^{(i-1)}_{\rm TS1,2}}]\notag\\
&\le \U{Pr}[y_i={\rm multi~click}|\bm{y}_{i-1},\bm{n^{(i-1)}_{\rm TS1,2}}].
\end{align}
This leads to
\begin{align}
\U{Pr}[n^{(i)}_{\rm TS1,2}\ge2|\bm{y}_{i-1},\bm{n^{(i-1)}_{\rm TS1,2}}]\le \frac{\U{Pr}[y_i={\rm multi~click}|
\bm{y}_{i-1},\bm{n^{(i-1)}_{\rm TS1,2}}]}
{\U{Pr}[y_i={\rm multi~click}|n^{(i)}_{\rm TS1,2}\ge2,\bm{y}_{i-1},\bm{n^{(i-1)}_{\rm TS1,2}}]}.
\label{app:LU}
\end{align}
We next consider lower-bounding 
$\U{Pr}[y_i={\rm multi~click}|n^{(i)}_{\rm TS1,2}\ge2,\bm{y}_{i-1},\bm{n^{(i-1)}_{\rm TS1,2}}]$. 
When $n^{(i)}_{\rm TS1,2}\ge2$, except for the case where all the $n^{(i)}_{\rm TS1,2}$ photons are 
contained in path 0 or path 1 at TS1 or TS2, the multi-click event always occurs. 
Here, we refer to the output paths of the Mach-Zehnder interferometer as ``path 0" for outputting bit 0 and ``path 1" for outputting bit 1 (see Fig.~\ref{fig:app}). 
Therefore, we only need to consider the case where all the photons are contained in either path 0 or path 1. 
For instance, if $n^{(i)}_{\rm TS1,2}\ge2$ photons exist on path 0 at TS1 or TS2, we have that the 
probability of obtaining the multi-click event is given as
$$
1-(1-\eta_3)^{n^{(i)}_{\rm TS1,2}}-\eta_3^{n^{(i)}_{\rm TS1,2}}\ge 
1-(1-\eta_3)^2-\eta_3^2=2\eta_3(1-\eta_3).
$$
Similarly, if $n^{(i)}_{\rm TS1,2}$ photons are present on path 1, their probability is
$2\eta_4(1-\eta_4)$. Using the assumption in Eq.~(\ref{assueta34}), we obtain 
$\U{Pr}[y_i={\rm multi~click}|n^{(i)}_{\rm TS1,2}\ge2,\bm{y}_{i-1},\bm{n^{(i-1)}_{\rm TS1,2}}]\ge 2\max\{\eta_3,\eta_4\}(1-\max\{\eta_3,\eta_4\})$, and hence 
Eq.~(\ref{app:LU}) results in
\begin{align}
\U{Pr}[n^{(i)}_{\rm TS1,2}\ge2|\bm{y}_{i-1},\bm{n^{(i-1)}_{\rm TS1,2}}]\le \frac{
\U{Pr}[y_i={\rm multi~click}|\bm{y}_{i-1},\bm{n^{(i-1)}_{\rm TS1,2}}]
}
{(1+2\max\{\delta^{(\rm BS)}_3,\delta^{(\rm BS)}_4\})(1/2-\max\{\delta^{(\rm BS)}_3,\delta^{(\rm BS)}_4\})}.
\end{align}
Taking the sum from 1 to $N_{\rm em}$ on both sides and applying Azuma's inequality~\cite{azumaineq}, 
the left-hand and right-hand sides approach $N_{\rm tag}$ and $N_{\rm multi}$ in the asymptotic limit, respectively. 
This ends the proof of Eq.~(\ref{l:keyrateTh}).

\end{document}